\documentclass[aps,prd,twocolumn,nofootinbib,showpacs,superscriptaddress]{revtex4-1}

\usepackage{amsfonts}
\usepackage{amsmath}
\usepackage{amssymb}
\usepackage{bm}
\usepackage{dcolumn}
\usepackage{epsfig}
\usepackage{graphicx}
\usepackage{graphics}
\usepackage[latin1]{inputenc}
\usepackage{latexsym}
\usepackage{rotating}
\usepackage{hyperref}
\usepackage[x11names]{xcolor}
\usepackage{xspace} % Sensible space treatment at end of simple macros
\usepackage{mathrsfs}
\usepackage{subfigure}
\usepackage{enumerate}
\usepackage{tabularx}
\usepackage{multirow,hhline}
\usepackage{booktabs}
\usepackage{siunitx}
\usepackage{array}
\newcolumntype{L}[1]{>{\raggedright\let\newline\\\arraybackslash\hspace{0pt}}m{#1}}
\newcolumntype{C}[1]{>{\centering\let\newline\\\arraybackslash\hspace{0pt}}m{#1}}
\newcolumntype{R}[1]{>{\raggedleft\let\newline\\\arraybackslash\hspace{0pt}}m{#1}}
\newlength{\Oldarrayrulewidth}

\hypersetup{
	colorlinks=true,
	linkcolor=magenta,
	filecolor=blue,
	citecolor = cyan,
	urlcolor=cyan,
}
\usepackage{ulem}
\definecolor{rufous}{rgb}{0.66, 0.11, 0.03}
\graphicspath{{../}}

\newcommand{\be}{\begin{equation}}
\newcommand{\ee}{\end{equation}}
\newcommand\ba{\begin{eqnarray}}
\newcommand\bse{\begin{subequations}}
\newcommand\ea{\end{eqnarray}}
\newcommand\ese{\end{subequations}}

\newcommand{\nn}{\nonumber}

\newcommand{\eq}{\,=\,}
\newcommand{\mg}[1]{\left|#1\right|}

\newcommand{\mat}{{\mbox{\tiny mat}}}

\newcommand{\DEF}{{\mbox{\tiny DEF}}}
\newcommand{\MO}{{\mbox{\tiny MO}}}

\newcommand{\jene}{\tilde{\epsilon}}
\newcommand{\jrho}{\tilde{\rho}}
\newcommand{\jpre}{\tilde{p}}
\newcommand{\rhomin}{\rho_{\min}}

\newcommand{\ppN}{{\mbox{\tiny PPN}}}

\newcommand{\crit}{{\mbox{\tiny crit}}}

\definecolor{red(ncs)}{rgb}{0.77, 0.01, 0.2}

\begin{document}

\title{
%Calculating scalar charges in massless scalar-tensor theories  of gravity}
Scalar Charges and Scaling Relations in Massless Scalar-Tensor Theories}

\author{David Anderson}
\affiliation{eXtreme Gravity Institute, Department of Physics, Montana State University, Bozeman, MT 59717, USA.}

\author{Nicol\'as Yunes}
\affiliation{eXtreme Gravity Institute, Department of Physics, Montana State University, Bozeman, MT 59717, USA.}

\date{\today}
%------------------------------------------------------------------------------------------------

%------------------------------------------------------------------------------------------------
\begin{abstract}\label{sec:abstract}
%------------------------------------------------------------------------------------------------ 

The timing of binary pulsars allows us to place some of the tightest constraints on modified theories of gravity. Perhaps some of the most interesting and well-motivated extensions to General Relativity are scalar-tensor theories, in which gravity is mediated by the metric tensor and a scalar field. These theories predict large deviations from General Relativity in the presence of neutron stars through a phenomenon known as scalarization. Neutron stars in scalar-tensor theories develop scalar charges, which directly enter the timing model for binary pulsars. In this paper, we calculate and tabulate these scalar charges in two popular, massless scalar tensor theories for a collection of neutron star equations of state that are compatible with constraints placed by the recent, gravitational wave observations of a binary neutron star coalescence. We then study these scalar charges and explore analytic scaling relations that allow us to predict their value in a large region of parameter space. Our results allow for the quick evaluation of the scalar charge in a large region of scalar-tensor theory parameter space, which has applications for gravitational wave tests of scalar-tensor theories, as well as binary pulsar experiments.

\end{abstract}
%------------------------------------------------------------------------------------------------

\maketitle
%\tableofcontents
\allowdisplaybreaks[4]
%------------------------------------------------------------------------------------------------
\section{Introduction}\label{sec:introduction}
%------------------------------------------------------------------------------------------------

%par1: We are now able to test gravity with extreme accuracy
Gravitational waves observations from aLIGO~\cite{Abbott:2016blz, Abbott:2016nmj, Abbott:2017vtc, Abbott:2017gyy, Abbott:2017oio, TheLIGOScientific:2017qsa} and the high precision timing of binary pulsars~\cite{Freire:2012mg, Kramer:2016kwa, Weisberg:2010zz, Wex:2014nva} has allow us to test Einstein's theory of General Relativity (GR) in the most extreme environments~\cite{Berti:2015itd, Yunes:2013dva}. With more NS-NS merger events, we will be able to tightly constrain the equation of state (EOS) and other properties of neutron stars (NS)~\cite{Abbott:2018exr, Abbott:2018wiz}. Furthermore, as radio astronomers continue to monitor binary pulsars systems, the errors in the timing model parameters will continue to decrease and help place tighter constraints on modified theories of gravity. Nonetheless, in order to investigate how these future observations will help us further test gravity, we first must understand the precise details of how observable predictions are modified. 

%par2: STTs are one of the most simple and well studied theories of gravity
A popular class of theories in the literature are scalar-tensor theories (STT) of gravity, in which gravity is not only mediated by the metric tensor but also by a long-range scalar field that is non-minimally coupled. Each theory in this class is defined by the choice of conformal coupling function, which mediates the degree of violation of the strong equivalence principle (SEP). Such theories were first studied by Jordan~\cite{Jordan-book, Jordan:1959eg}, Fierz~\cite{Fierz:1956zz}, Brans~\cite{Brans:1961sx}, and Dicke (JFBD) as the most natural alternatives to GR, and were later extended By Damour and Esposito-Far\'ese (DEF)~\cite{Damour:1992we, Damour:1993hw} to include higher order effects. A more recent extension of these theories was introduced by Mendes and Ortiz~\cite{Mendes:2016fby} (MO), which introduce a conformal coupling that replicates the behavior of including higher order scalar-field terms in the action~\cite{Damour:1996ke, Salgado:1998sg, Lima:2010na, Pani:2010vc}.

%par3: Binary pulsar syetsms present some of the best test beds for gravity
Solar system observations, like that of the perihelion shift of Mercury and of the Shapiro time delay, are able to place tight bounds on the parameters of STTs~\cite{Bertotti:2003rm, Will:2014kxa}. However, these are only measurements in the weak field regime where the gravitational potential is small. STTs are able to satisfy weak field constraints and still produce strong field deviations from GR through a phenomenon known as scalarization~\cite{Damour:1992we, Damour:1993hw, Barausse:2012da, Palenzuela:2013hsa}. Thus, one needs observations that probe the strong field, regions where non-linear effects like scalarization occur, in order to place tight constraints on STTs. While gravitational waves observations of binary NS coalescences with aLIGO and other detectors will be able to accomplish this regularly in the future, binary pulsar experiments are already able to probe and constrain STTs in the strong field to extremely high precision. By modeling the time of arrival (TOA) of pulses emitted from pulsar systems~\cite{DD1, DD2, Damour:1991rd}, one can take the observed data and place constraints on the underlying theory of gravity governing the motion of the binary. For this reason, binary pulsars are currently one of the best available testbeds for gravity. 

%par4: No one ahs done a full MCMC type calculation with bianry pulsar data to test GR
To perform any test, however, one must first know precisely how observables are modified in STTs, and this depends on the so-called scalar charges, i.e.~scalar quantities that determine how strongly a scalar field is sourced by an isolated neutron star. For example, the scalar charges enter directly into the parameterized-post-Keplerian (PPK) parameters used in testing STTs~\cite{Damour:1991rd, Damour:1996ke, Horbatsch:2011nh, Damour:2007uf, EspositoFarese:2004tu} with binary pulsars. To find these scalar charges, one must numerically obtain NS solutions in STTs, subject to certain physical boundary conditions at the core of the star and at spatial infinity. This numerical process is, at first sight, simple, yet in practice it need not be, mainly for the following two reasons. 

First, some of these charges can be numerically challenging to compute. The dominant scalar charge can indeed be read out easily from the leading 1/r piece of the scalar field at spatial infinity, given any numerical solution. Other scalar charges, however, depend on derivatives of the dominant charge with respect to the gravitational mass of the neutron star. During scalarization, the dominant scalar charge can change quite abruptly with respect to the asymptotic value of the scalar field, holding the baryonic mass of the star constant. This, in turn, leads to large spikes in the derivatives, which can be difficult to resolve if one is not careful. 

Second, tests of STTs requires knowledge of the scalar charges everywhere in parameter space, and this can be computationally costly. Markov-Chain Monte-Carlo (MCMC) methods that explore the posterior probability distribution of parameters requires the evaluation of the likelihood function hundreds of thousands to millions of times. Every evaluation requires the scalar charges, and if these have to be numerically computed every time the likelihood is needed, the MCMC exploration becomes computationally prohibitive. Clearly then, the MCMC exploration of parameter space in STTs would greatly benefit from a ``bank'' of calculated scalar charges.

In this paper, we study, calculate, tabulate, and analytically explore the three main scalar charges ($\alpha_A$, $\beta_A$, and $k_A$~\cite{Damour:1996ke}) that enter the PPK parameters of binary pulsar observations. We carry out this calculation both in the STT proposed by Damour-Esposito-Far\'ese~\cite{Damour:1992we, Damour:1993hw} and the one studied by Mendes-Ortiz~\cite{Mendes:2016fby} (hereafter referred to DEF theory and MO theory respectively). We explore a very large region of the parameter space spanned by the two coupling constants of these theories, using 11 different equations of state that are all consistent with neutron stars heavier than $2 M_{\odot}$, including a few that are also consistent with the recent gravitational wave observation of a neutron star coalescence~\cite{TheLIGOScientific:2017qsa}. 

Our main result is the construction of an accurate bank of scalar charges in these two theories that can now be used used in Bayesian model selection and parameter estimation studies of tests of STTs with binary pulsar observations. This bank is constructed both through direct numerical calculations, as well as through the exploration of certain analytic scaling relations. We determine the regime of parameter space in which the latter hold, and when they do, we use them to greatly accelerate the calculation of scalar charges in these regions of parameter space. The end result is a numerically accurate and dense bank of scalar charges that can be interpolated if necessary to provide charges everywhere in parameter space. 

%par5: While these calucaltions are simple to perfrom, they are too expensive to do on the fly, so we went through the effort to tabulate them
The remainder of this paper presents the details of the calculation summarized above. Section~\ref{sec:backgroun:field_equations} covers the basics of STTs along with observational constraints and includes a discussion of the scalar charges. Section~\ref{sec:NS} describes how NSs behave in STTs and begins to set up our numerical scheme to solve for the scalar charges. Section~\ref{sec:charges} details the calculations behind the scalar charges and what numerical techniques are needed to accurately explore the parameter space. Section~\ref{sec:data} provides a detailed description of our publicly available data files, along with instructions for how to use them and what their limitations are.   Section~\ref{sec:conclusion} concludes with a discussion of future work.

%------------------------------------------------------------------------------------------------

%------------------------------------------------------------------------------------------------
\section{Scalar-Tensor Theories}\label{sec:background}
%------------------------------------------------------------------------------------------------

%par1: Einstein frame action and definations
In this section, we introduce the basics of the class of STTs that we consider and establish the notation to be used in the rest of the paper. For completeness we present this class of theories from first principles using an action and provide a summary of the calculations needed to reach the field equations; we refer the reader to~\cite{Will:1993ns} for further details. We then describe the two STTs we study in this paper and present the current constraints on these theories. We conclude this section with a discussion of scalarization and the definition of scalar charges in these theories.

%------------------------------------------------------------------------------------------------
\subsection{Background and field equations}\label{sec:backgroun:field_equations}
%------------------------------------------------------------------------------------------------

In general, the massless STTs that we consider can be described in the \emph{Jordan frame} by an action of the form $\tilde{S} = \tilde{S}_{g} + \tilde{S}_\mat$, with the gravitational part taking the form
\be
	\tilde{S}_g \eq \int\dfrac{d^4x}{c} \dfrac{\sqrt{-\tilde{g}}}{4\kappa}\left[\phi \tilde{R} - \dfrac{\omega(\phi)}{\phi} \partial_\mu \phi \partial^\mu \phi\right]\,\,,
\label{eq:jordan-action}
\ee
where $\tilde{g}$ and $\tilde{R}$ are the determinant and Ricci scalar of the metric $\tilde{g}_{\mu\nu}$ respectively, $\phi$ is a scalar field, $\omega(\phi)$ is a coupling function of the scalar field, and $\kappa = 4 \pi G/c^4$ with $G$ the bare gravitational constant. The matter part of the action,  $\tilde{S}_\mat [\chi,\,\tilde{g}^{\mu\nu} ]$, is a functional of the matter fields $\chi$ that couple directly to the Jordan-frame metric. Therefore, the STTs we study in this paper are metric theories, and as such, laboratory clocks and rods measure time intervals and distances associated with $\tilde{g}_{\mu\nu}$.

While STTs can be completely described using the Jordan-frame action above, it is far more convenient to perform a conformal transformation that puts the action in a form reminiscent of the Einstein-Hilbert action. Let us then consider the transformation $\tilde{g}_{\mu\nu} = A(\varphi) g_{\mu\nu}$, where $g_{\mu\nu}$ is the \emph{Einstein-frame} metric\footnote{From this point on, we use an overhead tilde to represent quantities that are specifically in the Jordan frame. Quantities without overhead tildes should be assumed to be in the Einstein frame.}, so that the action becomes
\be
	S \eq \int \dfrac{d^4x}{c} \dfrac{\sqrt{-g}}{4\kappa}\left[ R - 2g^{\mu\nu} \partial_\mu \varphi \partial_\nu \varphi\right] + S_\mat\left[\chi, A^2(\varphi)g_{\mu\nu}\right]\,\,,
\label{eq:einstein-action}
\ee
where $g$ and $R$ are the now the determinant and Ricci scalar associated with the Einstein-frame metric $g_{\mu\nu}$. Notice that the matter fields now couple to $A^2(\varphi)g_{\mu\nu}$, and therefore, matter no longer falls along geodesics of the metric $g_{\mu\nu}$, but rather it is now also influenced by the scalar field $\varphi$.

The conformal transformation that takes Eq.~\eqref{eq:jordan-action} into Eq.~\eqref{eq:einstein-action} requires the conformal factor
\be
	A(\varphi) = \phi^{-1}\,\,,
\label{eq:conformal-transformation}
\ee
which then leads to a direct relation between $\varphi$ and $\phi$, given explicitly as
\be
	\alpha(\varphi)^2 \eq \left( \dfrac{d\,\ln A(\varphi)}{d \varphi}\right)^2 \eq \dfrac{1}{3 + 2\omega(\phi)}\,\,.
\label{eq:conformal-alpha}
\ee
One can think of $\alpha(\varphi)$ as the gradient of some ``conformal potential''~\cite{Anderson:2016aoi,Anderson:2017phb,Damour:1992kf,Damour:1993id}, defined by $V_{\alpha} \equiv \ln A(\varphi)$, and one can denote the ``curvature'' of this potential as $\beta(\varphi) = d\alpha/d\varphi$. The conformal potential is a simple way to understand and visualize the coupling between matter and the scalar field, which will be directly quantified by its slope and curvature in the field equations. Clearly then, the choice of $A(\varphi)$, or any of the above quantities for that matter, defines a particular member of this general class of STTs, which is ultimately a choice of exactly how the scalar field affects matter.

The variation of the Einstein-frame action with respect to the dynamical fields, $g_{\mu\nu}$ and $\varphi$, yields the field equations
\ba
	R_{\mu\nu} &\eq& 2 \partial_\mu \varphi \partial_\nu \varphi + 2\eta \left(T_{\mu\nu}^\mat - \dfrac{1}{2}g_{\mu\nu} T^\mat\right)\,\,,
	\label{eq:einstein-equation}\\
	\Box \varphi &\eq& -\kappa \alpha(\varphi)T^\mat\,\,,
	\label{eq:kg-equation}
\ea
where the Einstein frame stress-energy tensor is defined by
\be
	T_{\mu\nu}^\mat \equiv \dfrac{2c}{\sqrt{-g}}\left(\dfrac{\delta S_m}{\delta g^{\mu\nu}}\right)\,\,,
\label{eq:SET-definition}
\ee
and $T^\mat \equiv g_{\mu\nu} T_{\mu\nu}^\mat$ is its trace. The stress-energy tensor in the Einstein frame can be related to its Jordan-frame counterpart via the relation $T_{\mu\nu}^\mat = A^2(\varphi) \tilde{T}_{\mu\nu}^\mat$ by applying the conformal transformation to Eq.~(\ref{eq:SET-definition}). Assuming a perfect fluid description of the stress-energy tensor [see e.g.~Eq.~(\ref{eq:SET-perfect-fluid}) below] allows one to derive the relations $\epsilon = A^4 \jene$ and $p = A^4 \jpre$ between the energy density and pressure of the fluid in the different frames.
%------------------------------------------------------------------------------------------------

%------------------------------------------------------------------------------------------------
\subsection{Scalar-tensor models}\label{sec:background:models}
%------------------------------------------------------------------------------------------------

Scalar-tensor theories of the form described in the previous subsection allow for deviations from GR because of the new coupling that exists between matter and the scalar field. The particular choice of the conformal factor $A(\varphi)$, and likewise the formal coupling $\alpha(\varphi)$, defines the theory and plays a crucial role in understanding the observable modifications a theory predicts. The first model we consider is DEF theory and it is defined by
\ba
	A(\varphi) &\eq& e^{\beta_0 \varphi^2 /2}\,\,,
	\label{eq:DEF-conformal}\\
	V_{\alpha}(\varphi) &\eq& \dfrac{1}{2}\beta_0 \varphi^2\,\,,
	\label{eq:DEF-potential}\\
	\alpha(\varphi) &\eq& \beta_0 \varphi\,\,,
	\label{eq:DEF-coupling}\\
	\beta(\varphi) &\eq&  \beta_0\,\,,
	\label{eq:DEF-curvature}
\ea
where $\beta_0$ is a free coupling parameter. Aside from the JFBD theory in which $A(\varphi) = e^{\alpha_0 \varphi}$ , this is the simplest massless STT one can consider. One notices that the conformal potential is exactly a parabola whose curvature is precisely determined by the free parameter $\beta_0$. 

The other model we consider, which has gained attention in the past few years, is MO theory and it is defined by
\ba
	A(\varphi) &\eq& \left[\cosh\left(\sqrt{3}\, \beta_0 \varphi\right)\right]^{1/(3\beta_0)}\,\,,
	\label{eq:COSH-conformal}\\
	V_{\alpha}(\varphi) &\eq& \dfrac{1}{3\beta_0}\ln\left[\cosh\left(\sqrt{3} \,\beta_0 \varphi\right)\right]\,\,,
	\label{eq:COSH-potential}\\
	\alpha(\varphi) &\eq& \dfrac{\tanh\left(\sqrt{3}\,\beta_0 \varphi\right)}{\sqrt{3}}\,\,,
	\label{eq:COSH-coupling}\\
	\beta(\varphi) &\eq&  \beta_0 \,\text{sech}^2 \left( \sqrt{3}\,\beta_0 \varphi \right)\,\,,
	\label{eq:COSH-curvature}
\ea
where again $\beta_0$ is a free coupling parameter.  The MO theory was introduced as an analytic approximation to a more fundamental theory that includes quadratic terms of the scalar field coupled to curvature in the action~\cite{Damour:1996ke, Salgado:1998sg, Birrell:1982ix, Pani:2010vc, Mendes:2016fby}. This theory is functionally equivalent to DEF theory in the limit that $\varphi \rightarrow 0$, but it has strictly different behavior when the combination $\beta_0 \varphi \neq 0$. Therefore, these theories have distinctly different properties, and therefore, modify observables in different ways~\cite{Mendes:2016fby, Anderson:2017phb}.

%------------------------------------------------------------------------------------------------

%------------------------------------------------------------------------------------------------
\subsection{Solar System Constraints}\label{sec:background:constraints}
%------------------------------------------------------------------------------------------------

In principle, observations we make, whether they be in the solar system~\cite{Will:2014kxa, Bertotti:2003rm} or of binary pulsar systems~\cite{Kramer:2016kwa, Wex:2014nva, Damour:2007uf}, constrain the free parameters of the theory. Let us then consider how observables are modified in STTs. As an example, let us first consider the local value of Newton's gravitational constant. This quantity is given by
\be
	G_N \eq G\left[A^2_\infty\left(1+ \alpha^2_\infty\right)\right]\,\,.
\label{eq:G-newton}
\ee
where $G$ is the bare gravitational constant appearing in the action, and an $\infty$ subscript denotes quantities evaluated at $\varphi = \varphi_\infty$, e.g. $A_\infty \eq A(\varphi_\infty)$. The correction to the gravitational constant causes bodies to accelerate differently depending on the magnitude of scalar field, the parameters of the theory, and the bodies' composition through violations of the strong-equivalence principle. 

The choice of coupling parameter also determines the local value of the parameterized post-Newtonian (PPN) parameters~\cite{Nordtvedt:1970uv, Nordtvedt:1972zz, Will:1972zz}. Scalar-tensor theories are a class of fully conservative theories and therefore only pose modifications to the $\gamma_\ppN$ and $\beta_\ppN$ parameters~\cite{Will:1993ns, Will:2014kxa}. The former is given by
\be
	\mg{1 - \gamma_\ppN} \eq \dfrac{2 \alpha_\infty^2}{1 + \alpha_\infty^2}\,\,,
\label{eq:ppn-gamma}
\ee
while the latter is given by
\be
	\mg{1 - \beta_\ppN} \eq \dfrac{\beta_\infty\alpha_\infty^2}{2\left(1 + \alpha_\infty^2\right)^2}\,\,,
\label{eq:ppn-beta}
\ee
where as before $\alpha_{\infty} = \alpha(\varphi_{\infty})$ and $\beta_{\infty} = \beta(\varphi_{\infty})$. The $\gamma_{\ppN}$ parameter is a measure of the spatial curvature induced by a unit rest mass, while the $\beta_{\ppN}$ parameter is a measure of the amount of non-linearity in the superposition law for gravity. The $\gamma_{\ppN}$ parameter has been measured from the Shapiro time delay observed by the Cassini spacecraft~\cite{Bertotti:2003rm, Will:2014kxa}, and it is constrained to $\mg{1 - \gamma_\ppN} \lesssim 2.3 \times 10^{-5}$. The $\beta_{\ppN}$ parameter is measured from observations of the perihelion shift of Mercury~\cite{Will:2014kxa}, and it is constrained to  $\mg{1 - \beta_\ppN} \lesssim 8\times 10^{-5}$.

The notation we have used above is slightly different from what is found in the literature so let us clarify this here. Typically, instead of using $\alpha_\infty$ and $\beta_{\infty}$, some papers that studied DEF theory used a different set of parameters $\{ \alpha_0, \, \beta_0\}$. This is because if one modifies Eq.~\eqref{eq:DEF-coupling} in DEF theory to 
\be
	\alpha(\varphi) \eq \alpha_{0} + \beta_0 \varphi\,\,,
	\label{eq:DEF-conformal-new}
\ee
and sets $\varphi_{\infty} = 0$, then $\alpha(\varphi_{\infty}) = \alpha_{0}$ and $ \beta(\varphi_{\infty})=\beta_{0}$, and all observables can be entirely parameterized by the set $\{ \alpha_{0},\beta_{0} \}$. This parameterization is identical to our description of DEF theory appearing in Eq.~\eqref{eq:DEF-conformal}, provided that one enforces $\varphi_\infty = \alpha_0/\beta_0$~\cite{Damour:2007uf}, which is the choice we make in this paper. When considering MO theory, however, $\alpha(\varphi_{\infty}) \neq \alpha_{0} = \varphi_{\infty} \beta_{0}$ and $ \beta(\varphi_{\infty}) \neq \beta_{0}$, as one can easily see from Eqs.~\eqref{eq:COSH-conformal}-\eqref{eq:COSH-curvature}.

In this paper, we want both theories to share the same free parameters $\{\alpha_0,\,\beta_0\}$ and, therefore, the quantities that enter the PPN parameters, $\{ \alpha_\infty,\,\beta_\infty\}$, are different functions of $\{\alpha_0,\,\beta_0\}$ in the two theories. These functions are 
\ba
	\alpha_\infty^{\DEF} &\eq& \alpha_0\,\,,
	\label{eq:alpha_bar-DEF}\\
	\beta_\infty^{\DEF} &\eq& \beta_0\,\,,
	\label{eq:beta_bar-DEF}
\ea
in DEF theory, and
\ba
	\alpha_\infty^{\MO} &\eq& \tanh\left( \sqrt{3}\,\alpha_0\right)/\sqrt{3}\,\,,
	\label{eq:alpha_bar-MO}\\
	\beta_\infty^{\MO} &\eq& \beta_0\,\text{sech}^2 \left( \sqrt{3}\,\alpha_0\right)\,\,.
	\label{eq:beta_bar-MO}
\ea
in MO theory. These choices have the advantage of reducing $(\alpha_{\infty},\beta_{\infty})$ to the known relations of DEF theory, while properly generalizing them to MO theory. 

%------------------------------------------------------------------------------------------------

%------------------------------------------------------------------------------------------------
\subsection{Scalarization and binary pulsar constraints}\label{sec:background:scalarization}
%------------------------------------------------------------------------------------------------

Solar system observations have the ability to place tight constraints on STTs through weak field observations~\cite{Will:2014kxa}. STTs, however, are able to satisfy these constraints and still deviate substantially from GR inside and near NSs~\cite{Damour:1992we, Damour:1993hw, Barausse:2012da, Palenzuela:2013hsa}. The strong field deviations are caused by a phenomenon known as scalarization, in which the scalar field can grow rapidly towards order unity inside NSs even when the asymptotic value, that which is constrained by Solar System observations, approaches zero. 

The key behind this rapid growth is the existence of an instability in the field equations when a star reaches a sufficiently large compactness. The onset of this instability is analogous to spontaneous magnetization in ferromagnets~\cite{Damour:1996ke}. To understand this, consider the external scalar field far from a neutron star, labeled $A$,  $\varphi \eq \varphi_\infty + G \, \omega_A/r + \mathcal{O}(1/r^2)$, where $\omega_A$ is a type of ``charge'' that is energetically conjugate to the external scalar field, 
\be
	\omega_A \eq -\dfrac{\partial m_A}{\partial \varphi_\infty}\,\,,
\label{eq:scalar_charge-definition}
\ee
with $m_A$ the total gravitational mass of the NS. When one considers a sequence of neutron stars of masses $m_A$, $\omega_A$ can become suddenly non-zero at a critical value of the mass or compactness of the star. This sudden activation of the scalar field is what is referred to as spontaneous scalarization. When a NS is scalarized, $\omega_A \neq 0$ and the scalar field is excited above its background value, leading to local gravitational effects that will generically be different than those in GR. 

%, such as orbital decay and periastron advance, in binary system containing NSs are inherently different from the predictions of GR. 

For binary pulsar tests, it is convenient to introduce certain quantities that enter the PPK parameters of the binary pulsar timing model. We call these parameters \emph{scalar charges} in this paper, the first of which is defined by
\be
	\alpha_A \eq \left.-\dfrac{\omega_A}{m_A} \eq \dfrac{\partial \ln m_A}{\partial \varphi_\infty}\right|_{\bar{m}_A}\,\,,
\label{eq:charge_1}
\ee
which plays the role of an effective coupling between the scalar field and the $A$th NS in the binary. This quantity is the strong field counterpart of the $\alpha_\infty$ parameter introduced in the previous section. Similarly, there is a strong field counterpart to the $\beta_\infty$ parameter of the previous section, namely
\be
	\beta_A \eq \left.\dfrac{\partial \alpha_A}{\partial \varphi_\infty}\right|_{\bar{m}_A}\,\,,
\label{eq:charge_2}
\ee
which encodes higher order effects associated with the exchange of multiple scalar particles between the binary components. Lastly, there is one more charge that enters the PPK parameters, namely
\be
	k_A \eq \left.\dfrac{\partial \ln I_A}{\partial \varphi_\infty}\right|_{\bar{m}_A}\,\,,
\label{eq:charge_3}
\ee
where $I_A$ is the moment of inertia of the NS. Similarly to how $\alpha_A$ was an effective coupling between the mass of the NS and the scalar field, this quantity acts as a coupling between the field and the star's spin angular momentum, and it describes how the NS's inertia reacts to the presence of an external scalar field. All of these scalar charges must be calculated while holding the baryonic mass of the star constant, as they measure the ``sensitivity'' of a star to the external scalar field. The main goal of this work is to calculate these scalar charges and make them publicly available. As such, we will provide the details of these calculations later in \S\ref{sec:charges} after we introduce the relevant framework needed to understand NSs in STTs.

%------------------------------------------------------------------------------------------------

%------------------------------------------------------------------------------------------------
\section{Compact Stars in Scalar-Tensor Gravity}\label{sec:NS}
%------------------------------------------------------------------------------------------------

%par00: some intro to the importance and topic of the section
In this section we discuss how compact stars behave in STTs and introduce some of the basics of scalarization from an analytic perspective. We focus our attention on isolated, slowly-rotating stars because the binary pulsar systems that we wish to provide scalar charges for are widely separated. We begin with a discussion of the interior spacetime of slowly-rotating compact stars and then discuss the exterior spacetime and how one connects it to the interior. Following the discussion of the field equations, we discuss the different types of equations of state one can use and conclude with an analytic discussion of scalarization in the different regions of parameter space.

%par1: Introduction to the Hartle slow rotation metric
Before we begin, however, let us discuss how we will describe compact stars in STTs. We focus on a stationary, axisymmetric spacetime, which allows for a description of a star that is slowly rotating. Following closely the work of~\cite{Damour:1996ke}, we make the metric ansatz proposed by Hartle~\cite{Hartle:1967he} 
\ba
	ds^2 &\eq& g_{\sigma \delta} \; dx^\sigma dx^\delta \eq -e^{\nu(\rho)}c^2 dt^2 + e^{\lambda(\rho)} d\rho^2 + \rho^2 d\theta^2 \,\,\nn\\
	&\,&+ \, \rho^2\sin^2\theta \left(d\phi + \left[\omega(\rho,\theta) - \Omega\right]dt\right)^2\,\,,
\label{eq:line-element}
\ea
in which one only keeps terms to first order in the star's angular velocity $\Omega= u_\phi / u_t$, where $u^\mu$ is the fluid's four velocity. In this ansatz, the metric functions $(\nu,\lambda)$ are zeroth-order in rotation and functions of the radial coordinate $\rho$ only, while the metric function $\omega$ is first-order in rotation and depends both on $\rho$ and the polar angle $\theta$.  For practical and physical convenience, we also redefine the $g_{rr}$ component of the metric through the interior mass function $\mu(\rho)$ defined via
\be
	e^{\lambda(\rho)} \eq \left(1 - \dfrac{2 \mu(\rho)}{\rho}\right)^{-1}\,\,.
\label{eq:metric_rr}
\ee
Moreover, we model the NS matter with a perfect fluid stress-energy tensor given by
\be
	T_{\mu\nu} \eq (\epsilon + p)u_\mu u_\nu + p g_{\mu\nu}\,\,,
\label{eq:SET-perfect-fluid}
\ee
where $\epsilon$ is the fluid's energy density and $p$ is the pressure, both in the Jordan frame.

%------------------------------------------------------------------------------------------------

%------------------------------------------------------------------------------------------------
\subsection{Interior Spacetime}\label{sec:NS:interior}
%------------------------------------------------------------------------------------------------

The scalar-tensor field equations with this metric ansatz are similar to those in GR. At zeroth-order in rotation, the field equations for the diagonal components of the metric are identical to those in the spherical case, which have already been studied in detail in the literature~\cite{Damour:1996ke}. At first order in rotation, one finds a second-order equation for the $\omega$ metric function, which can be converted into an ordinary, second-order differential equation in $\rho$ through a Legendre decomposition in $\theta$~\cite{Damour:1996ke}. Such a decomposition reveals that only the $\ell=1$ mode in the Legendre decomposition has support. In particular, the field equations can be written in the first-order form~\cite{Damour:1996ke}
\bse
\be
	\mu' \eq \kappa \rho^2 A^4(\varphi) \jene + \dfrac{1}{2}\rho(\rho - 2 \mu)\psi^2\,\,,
\label{eq:mu}
\ee
\be
	\nu' \eq 2\kappa\dfrac{\rho^2 A^4(\varphi) \jpre}{\rho - 2 \mu}+ \rho \psi^2+ \dfrac{2\mu}{\rho(\rho - 2 \mu }\,\,,
\label{eq:nu}
\ee
\be
	\varphi' \eq \psi\,\,,
\label{eq:phi}
\ee
\ba
	\psi' &\eq &\kappa \dfrac{\rho A^4(\varphi)}{\rho - 2 \mu}\left[\alpha(\varphi) (\jene - 3 \jpre) + \rho \psi (\jene - \jpre) \right]\,\,\nn\\
	&\,& - \dfrac{2(\rho - \mu)}{\rho(\rho - 2 \mu)} \psi\,\,,
\label{eq:psi}
\ea
\be
	\jpre' \eq -(\jene + \jpre) \left[ \dfrac{\nu'}{2} + \alpha(\varphi)\psi \right]\,\,,
\label{eq:p}
\ee
\be
	\bar{m}' \eq 4 \pi G \jrho A^3(\varphi)\dfrac{\rho^2}{\sqrt{1 - 2 \mu / \rho}}\,\,,
\label{eq:mass}
\ee
\be
	\omega' \eq \varpi\,\,,
\label{eq:omega}
\ee
\ba
	\varpi' &\eq& \kappa\dfrac{\rho^2 }{\rho - 2 \mu}A^4(\varphi)(\jene+\jpre)\left(\varphi + \dfrac{4 \omega}{\rho}\right) \,\,\nn\\
	&\,& + \left(\rho\psi^2 - \dfrac{4}{\rho}\right)\varpi\,\,,
\label{eq:varpi}
\ea
\label{eq:main-equation}
\ese
where primes denote derivatives with respect to the radial coordinate $\rho$. Notice that we have also included an equation for the enclosed baryonic mass of the star $\bar{m}(\rho)$, which gives the total baryonic mass through $\bar{m}_A = \bar{m}(R) = \int \jrho \, u^t \sqrt{-\tilde{g}} \, d^3x$, where $R$ marks the surface of the star, where by definition the pressure vanishes. In Eq.~(\ref{eq:main-equation}) we have explicitly used the Jordan-frame fluid variables $\jene$, $\jrho$,\,and $\jpre$ since these are the physical quantities that are measured by observations and appear in the EOSs that are discussed below.

From a numerical standpoint these equations pose a problem near the center of the star at $\rho=0$ since the equations diverge there. The proper way to deal with this is to expand the equations about $\rho=0$ and start the numerical integration at some arbitrary small distance away from the center, say $\rho_{\min}$. Expanding Eqs.~(\ref{eq:main-equation}) and evaluating at $\rho_{\min}$ gives the boundary conditions
\bse
	\be
		\mu(\rhomin) \eq 0\,\,,
	\label{eq:cBC-mu}
	\ee
	\be
		\nu(\rhomin) \eq 0\,\,,
	\label{eq:cBC-nu}
	\ee
	\be
		\varphi(\rhomin) \eq \varphi_c\,\,,
	\label{eq:cBC-phi}
	\ee
	\be
		\psi(\rhomin) \eq \left( \dfrac{\rhomin}{3}  \right)  \eta A^4(\varphi_c) \alpha(\varphi_c) \left[\jene_c - 3 \jpre_c\right]\,\,,
	\label{eq:cBC-psi}
	\ee
	\be
		\jrho(\rhomin) \eq \jrho_c\,\,,
	\label{eq:cBC-rho}
	\ee
	\be
		\bar{m}(\rhomin) \eq 0\,\,,
	\label{eq:cBC-mass}
	\ee
	\be
		\omega(\rhomin) \eq 1\,\,,
	\label{eq:cBC-omega}
	\ee
	\be
		\varpi(\rhomin) \eq \left(\dfrac{4}{5}\rhomin\right) \eta A^4(\varphi_c) \alpha(\varphi_c) \left[\jene_c - 3 \jpre_c\right]\,\,,
	\label{eq:cBC-varphi}
	\ee
\label{eq:central-conditions}
\ese
where $\jene_c = \jene(\jrho_c)$ and $\jpre_c = \jpre(\jrho_c)$ are the central values of the Jordan-frame energy density and pressure respectively and are defined by the EOS. The values of $\varphi_c$ and $\jrho_c$ are chosen independently and define a particular solution to the field equations. Equations~(\ref{eq:main-equation}) and (\ref{eq:central-conditions}) allow one to integrate the field equations from $\rho = \rhomin$ to any arbitrary radius, even outside the star as the field equations describe the entire spacetime. A more computationally efficient method, however, is to use an analytic solution in vacuum that is valid in the exterior of the star, and then, match it to the interior solution at $\rho=R$, as we describe in the next section.
%------------------------------------------------------------------------------------------------

%------------------------------------------------------------------------------------------------
\subsection{Exterior spacetime}\label{sec:NS:exterior}
%------------------------------------------------------------------------------------------------

%par4: Introduction to the Just spacetime and exterior matching conditions
The exterior solution to the field equations [Eqs.\eqref{eq:main-equation}] were found by Just in the late 1950s~\cite{Just:1959}. In the coordinates introduced by Just, the exterior metric takes the form
%\footnote{As mentioned in Ref.~\cite{Damour:1996ke}, this results from a boundary condition upon demanding that $\omega = \Omega$ as $\rho\rightarrow \infty$. This is an arbitrary choice and does not change the physics, but moreover it allows the cross terms in the axisymmetric spacetime to vanish far from the star. }
%
%
\be
	ds^2 \eq -e^{\nu} c^2 dt^2 + e^{-\nu}\left[dr^2 + (r^2 - ar)(d\theta^2 + \sin^2\theta\,d\phi^2)\right]\,\,,
\label{eq:line-element-Just}
\ee
where one has
\be
	e^{\nu(r)} \eq \left(1 - \dfrac{a}{r}\right)^{b/a}\,\,,
\label{eq:Just-tt}
\ee
while the scalar field takes the form
\be
	\varphi(r) \eq \varphi_\infty + \dfrac{d}{a}\,\ln \left(1 - \dfrac{a}{r}\right)\,\,,
\label{eq:Just-phi}
\ee
where $a$, $b$, and $d$ are integration constants and are constrained by the relation $a^2 - b^2 = 4d^2$. The integration constants can all be expressed in terms of the gravitational mass of the star $m_A$ and some effective coupling constant $\alpha_A$, which in this context plays the role of the scalar charge in Eq.~(\ref{eq:charge_1}). These relations take the form
\bse
	\be
		b = 2 \dfrac{G}{c^2} m_A\,\,,
	\ee
	\be
		\dfrac{a}{b} \eq \sqrt{1+ \alpha_A}\,\,,
	\ee
	\be
		\dfrac{d}{b} \eq \dfrac{1}{2}\alpha_A\,\,.
	\ee
\label{eq:Just-constants}
\ese

The coordinates used in the Just spacetime above are not the same as those used in the Hartle ansatz of the interior metric.  Comparing the two line elements, one finds that
\bse
	\be
		\rho \eq r\left(1 - \dfrac{a}{r}\right)^\frac{a-b}{2a}\,\,,
	\ee
	\be
		e^{\lambda(\rho)} \eq \left(1 - \dfrac{a}{r}\right) \left( 1 -\dfrac{a+b}{2r}\right)^{-2}\,\,.
	\ee
\label{eq:Just-relatiom}
\ese
Given these relations, we can now read off the total gravitational mass of the star $m_A$ from the $1/\rho$ behavior of $g_{tt}$, or $g_{\rho\rho}$, and the star's $z$-component of the total angular momentum $J_A$ from the $1/\rho^2$ portion of $g_{t\phi}$. We can recast the later in terms of the $1/\rho^3$ behavior of $\omega$ as
\be
	\omega \eq \Omega - \dfrac{G}{c^2}\dfrac{2 J_A}{\rho^3} + \mathcal{O}\left(\rho^{-4}\right)\,\,,
\label{eq:ang-velocity}
\ee
in which case the moment of inertia follows as 
\be
	I_A \eq \dfrac{J_A}{\Omega}\,\,.
\label{eq:moment-inertia}
\ee

By directly integrating the equation for $\omega(\rho)$ and inserting Eqs.~(\ref{eq:Just-relatiom}) into Eqs.~(\ref{eq:main-equation}), we arrive at a set of relations that are valid at the stellar surface~\cite{Damour:1996ke} (with a subscript $s$ denoting values at the surface of the star), namely
\bse
	\be
		R \equiv \rho_s\,\,,
	\label{eq:BC-r}
	\ee
	\be
		\nu'_s \equiv R \psi_s^2 + \dfrac{2 \mu_s}{R(R - 2 \mu_s)}\,\,,
	\label{eq:BC-nuprime}
	\ee
	\be
		\alpha_A \equiv \dfrac{2 \psi_s}{\nu'_s}\,\,,
	\label{eq:BC-charge}
	\ee
	\be
		Q_1 \equiv \sqrt{1+ \alpha_A^2}\,\,,
	\label{eq:BC-Q1}
	\ee
	\be
		Q_2 \equiv \sqrt{1 - 2 \mu_s / R}\,\,, 
	\label{eq:BC-Q2}
	\ee
	\be
		\hat{\nu}_s \equiv - \dfrac{2}{Q_1} \tanh^{\mbox{\footnotesize -1}} \left( \dfrac{Q_1}{1 + 2/(R \nu'_s)}\right)\,\,,
	\label{eq:BC-nuhat}
	\ee
	\be
		\varphi_\infty \equiv \varphi_s - \dfrac{1}{2}\alpha_A \hat{\nu}_s\,\,,
	\label{eq:BC-phiinf}
	\ee
	\be
		\dfrac{G}{c^2}m_A \equiv \dfrac{1}{2} \nu'_s R^2 Q_2\,\exp\left(\dfrac{1}{2} \hat{\nu}_s\right)\,\,,
	\label{eq:BC-massADM}
	\ee
	\be
		\dfrac{G}{c^2} J_A \equiv \dfrac{1}{6}\varpi_s R^4 Q_2 \,\exp\left(-\dfrac{1}{2} \hat{\nu}_s\right)\,\,,
	\label{eq:BC-J}
	\ee
	\ba
		\Omega &\equiv& \omega_s - \dfrac{c^2}{G^2}\dfrac{3 J_A}{4 m_A^3 (3 - \alpha_A^2)}\left\lbrace  e^{2\hat{\nu}_s} - 1 + \dfrac{4 G m_A}{R c^2}e^{\hat{\nu}_s}\,\right.\,\,\nn\\
		&\,& \left. \times \left[ \dfrac{2 G m_A}{R c^2} + e^{\hat{\nu}_s/2} \cosh \left( \dfrac{1}{2}Q_1 \hat{\nu}_s \right) \right] \right\rbrace \,\,,
	\label{eq:BC-Omega}
	\ea
\label{eq:surface-boundary}
\ese
These set of relations allow us to extract important observables at $\rho= \infty$ by simply knowing their appropriate values at the surface of the star.

%------------------------------------------------------------------------------------------------

%------------------------------------------------------------------------------------------------
\subsection{Equations of State}\label{sec:NS:eos}
%------------------------------------------------------------------------------------------------

%par1: Explain importance of EOS and why we look at multiple ones
The description of the problem is not complete without first knowing how the fluid properties $\{\jrho,\,\jpre,\,\jene\}$ depend on one another. This is accomplished by an EOS and it allows us to close the system of equations. In this subection we describe the three types of EOSs considered in this paper in detail.

%par2: Maybe have a big figure of all EOS mr curves and some stuff like that

%------------------------------------------------------------------------------------------------

%------------------------------------------------------------------------------------------------
\subsubsection{Polytropes}\label{sec:NS:eos:polytrope}
%------------------------------------------------------------------------------------------------

%par1: For comparison to DEF 96 we first present a polytopic equations of state...
The most simple EOS to consider for NSs is a polytropic equation of state in which the the pressure and baryonic density are related through a power law, i.e.
\be
	\jpre \eq K\jrho_0 \left(\dfrac{\jrho}{\jrho_0}\right)^\Gamma\,\,,
\label{eq:polytrope}
\ee
where $K$ is the polytropic constant and $\Gamma$ is the adiabatic index of the fluid. A particular choice for $K$ and $\Gamma$ define the EOS as well as some macroscopic properties of the NS, such as the maximum mass and compactness. In~\cite{Damour:1996ke}, a polytropic EOS was used in the calculation of the ``gravitational form factors'' (what we call the scalar charges in this paper) and therefore we will use the same polytropic EOS here to validate our computational algorithm. In particular, we will choose
\be
	\Gamma \eq 2.34 \,\,\,\,\,,\,\,\,\,\, K \eq 0.0195\,\,,
\label{eq:polytropic-parameters}
\ee
with a fiducial baryonic mass density $\jrho_0 = 1.66 \times 10^{14}$ g cm$^{-3}$, to make comparisons to the results in that paper, which we present later in Fig.~\ref{fig:def96_comp}.

The baryonic mass density only appears in the integral for calculating the baryonic mass of the star, and therefore, we need another relation relating pressure and baryonic density to the total energy density. The first law of thermodynamics provides such a relation:
\be
	\jene \eq \jrho + \dfrac{\jpre}{\Gamma - 1}\,\,.
\label{eq:first-thermo}
\ee
Equations~(\ref{eq:polytropic-parameters}) and (\ref{eq:first-thermo}) can now be inserted directly into Eqs.~(\ref{eq:main-equation}) to provide a complete description of the interior of the NS.

While polytopes are very simple analytic EOSs that facilitate quick numerical calculations, they are an oversimplification of the true microphysics occurring inside the NS. Polytropes assume the same functional dependence between pressure and density throughout the entire star and do not account for any real differences that exist in different density regimes. Due to this reason, we only use polytopes as a proof of concept for the existence of scalarization and to make valid comparison between our results and those presented in Ref.~\cite{Damour:1996ke} to ensure numerical consistency.
%------------------------------------------------------------------------------------------------

%------------------------------------------------------------------------------------------------
\subsubsection{Tabulated Equations of state}\label{sec:NS:eos:tabulated}
%------------------------------------------------------------------------------------------------

%par1: Talk about full eos and tabulated forms
A complete description of the microphysics occurring inside NSs requires the full modeling of $N$-body quantum systems at extremely high pressures and densities. The calculations required to solve for these relations is very expensive and not practical on the fly for every density and pressure inside a NS. Moreover, because NS type densities cannot be observed in laboratories on Earth there is uncertainty on what physics is actually taking place at these densities. There have been numerous models proposed to describe matter at supra-nuclear densities and they have been tabulated such that one can interpolate them as needed. 

For the purpose of this paper, we consider a wide range of tabulated EOSs that produce NSs with masses that are consistent with observations, most notably that of a near $2 M_\odot$ pulsar in J0348+0432. We consider 11 different tabulated EOS~\cite{Lattimer:2000nx} that satisfy this constraint: AP3-4~\cite{Akmal:1998cf}, ENG~\cite{Engvik:1995gn}, H4~\cite{Lackey:2005tk}, MPA1~\cite{Muther:1987xaa}, MS0~\cite{Mueller:1996pm}, MS2~\cite{Mueller:1996pm}, PAL1~\cite{Prakash:1988md}, SLy4~\cite{Douchin:2001sv}
\footnote{The SLy4 EOS we use here is commonly denoted as simply SLy in the literature.}
, and WFF1-2~\cite{Wiringa:1988tp}. All but one (H4) of these EOSs contain plain nuclear matter and \emph{do not} contain any form of strange matter. Because these EOS arise from numerical calculations that include true microphysics (or at least various justified approximations) we consider these to be the most physically relevant of the EOSs that we consider and are the ones we use for our main results. More importantly, however, many of these EOSs are consistent with aLIGO's constraints placed from the observation of coalescing NSs~\cite{Abbott:2018exr}.

%par2-3: introduce the various EOSs and appropriate details (see Read et al. PP paper) and refences

%------------------------------------------------------------------------------------------------

%------------------------------------------------------------------------------------------------
\subsubsection{Piecewise Polytropes}\label{sec:NS:eos:piecewise}
%------------------------------------------------------------------------------------------------

A useful compromise between the simple polytropic and tabulated EOSs is a piecewise polytropic model that stitches together multiple polytropes in different density regions inside the NS. In particular, we consider the piecewise polytropic EOSs studied in Ref.~\cite{Read:2008iy} in which the authors developed a parameterized model that can accurately capture the feature of tabulated EOS. Of the 34 EOSs that the authors fit their model to, 8 of them overlap with the set of tabulated EOSs that we consider in this paper\footnote{One might notice that, aside from the PAL1 EOS that is explicitly not contained in Ref.~\cite{Read:2008iy}, there are 10 EOSs listed in the previous section that correspond to the EOSs that were fit in this paper. The data we received from Norbert Wex came from the original work by Lattimer and Prakash~\cite{Lattimer:2000nx} and in fact the EOSs labeled MS0 and MS2 are \emph{different} from those of the same name appearing in Ref.~\cite{Read:2008iy}. There appears to be a mismatch in nomenclature that we feel is worth pointing out.}. While these approximations have been used extensively in the literature for their convenience, we later investigate how these approximations affect the scalar charges that are developed in NSs. Hence, let us briefly discuss these EOSs, referring to Ref.~\cite{Read:2008iy} for a more detailed and complete description.

Similarly to the standard polytropic EOS, the various regions inside of the NS are described by a polytrope of the form
\be
	\jpre \eq K_i\,\jrho^{\Gamma_i}\,\,,
\label{eq:piecewise-polytrope}
\ee
where now $\Gamma_i$ is the adiabatic index for the $i$th region of the NS and $K_i$ is the polytopic constant chosen to ensure continuity at the boundaries between regions. Similarly to the single polytrope case, the first law of thermodynamics in Eq.~(\ref{eq:first-thermo}) is used to find the energy density in each region
\be
	\jene_i \eq (1+a_i)\jrho_i + \dfrac{1}{\Gamma_i - 1} K_i \jrho_i^{\Gamma_i}\,\,,
\label{eq:pp_first_thermo}
\ee
where 
\be
	a_i \eq \dfrac{\jene(\jrho_{i-1})}{\jrho_{i-1}} -1 -\dfrac{K_i}{\Gamma_i - 1}\jrho_{i-1}^{\Gamma_i -1}\,\,,
\label{eq:pp-integration-constants}
\ee
is an integration constant that must be present in order to ensure that all fluid variable are continuous across the boundaries between regions. For the single polytrope case, the requirement that $\jene/\jrho =1$ in the limit that $\jrho\rightarrow 0$ forces $a=0$ and thus reduces Eq.~(\ref{eq:pp_first_thermo}) to Eq.~(\ref{eq:first-thermo}).

We adopt a low-density EOS for the crust of the NS that is identical to the one presented in Table II of Ref.~\cite{Read:2008iy} where the SLy (SLy4 as appearing in this paper) is independently fit for $\jrho \lesssim 10^{12}$ g/cm$^3$. The boundary between the crust and high-density EOS is then determined by the intersection of the respective polytropes and is ultimately determined by the value of $\Gamma_1$. Then, at a fixed baryonic density of $\jrho = 10^{14.7}$ g/cm$^3$ and best fit pressure $\jpre_1 = \jpre(\jrho_i)$ the first region is matched to a second region with adiabatic index of $\Gamma_2$ and polytropic constant $K_2$. Another match is then performed at another boundary $\jrho = 10^{15}$ g/cm$^3$ to an even higher density region with constants $\Gamma_3$ and $K_3$. The complete set of parameters $\{\log(\jpre),\,\Gamma_1,\,\Gamma_2,\,\Gamma_3\}$ represents the best fit values found in Table III of Ref.~\cite{Read:2008iy}. For our purposes, we focus exclusively on the approximations of AP3 and SLy4 to compare scalar charges between tabulated and piecewise polytropic EOSs.
%------------------------------------------------------------------------------------------------

%------------------------------------------------------------------------------------------------
\section{Calculating the Scalar Charges }\label{sec:charges}
%------------------------------------------------------------------------------------------------

%par1: Introduction to the section and overall breakdown
Now that we have a full description of the problem at hand, we solve the equations numerically to obtain the relevant scalar charges appearing in Eqs~(\ref{eq:charge_1})-(\ref{eq:charge_3}). We begin this section with an analytic description of the scalar charges, and then proceed with a description of our numerical methods for solving the field equations and extracting the various scalar charges. We then recap, but in more detail, the different regions of parameter space that we are concerned with and discuss their importance. We first present a comparison between some of our results and those originally found in Ref.~\cite{Damour:1996ke}, and then discuss our full results in each region of parameter space.

%------------------------------------------------------------------------------------------------
\subsection{Analytic investigation of scalarization}\label{sec:NS:analytic}
%------------------------------------------------------------------------------------------------

% the basic example of how scalarization can arise in the negative beta case
The phenomenon of scalarization has been well studied in the literature over the past decades, particularly in the context of spontaneous scalarization occurring when $\beta_0 \lesssim -4.3$. However, it is useful to review what happens outside of this regime as well since there are still non-linear effects coming into play, particularly when $\beta_0 \gtrsim -3.5$. In this subsection we review the analytics that help guide our calculations in the different regions of parameter space.

Scalarization can be understood from an analytic standpoint when one makes a few simple approximations. In both theories we consider here, the conformal coupling takes the form $\alpha(\varphi) = \beta_0 \varphi + \mathcal{O}(\varphi^2)$ in the limit that $\beta_0 \varphi$ is relatively small compared to unity. Let us now consider the field equation for the scalar field in Eq.~(\ref{eq:kg-equation}), but instead of considering the full nonlinear equation, we make a weak field approximation such that $\Box \rightarrow \delta_{ij} \nabla_i \nabla_j \rightarrow \nabla_r^2$ with the last term being the radial portion of the flat space Laplacian in spherical coordinates. We assume that $T_\mat$ is constant since we are considering weakly gravitating systems. As mentioned in~\cite{Damour:1993hw}, we do not expect the trace of the stress-energy tensor to be negative for weakly gravitating systems, but it is fruitful to leave its sign general and consider the full breadth of parameter space with the same analytics. 

The assumptions made thus far allow one to write the equation of motion for the scalar field as
\be
	\nabla_r^2 \varphi \eq - C^2\varphi\,\text{sign}(\beta_0\,T_\mat)\,\,,
\label{eq:analytic-DEF}
\ee
where we have introduced the constant $C^2 = \kappa|\beta_0\,T_\mat|$ for $r<R$, which vanishes when $r>R$. The field equation is still subject to the same boundary conditions as before, and thus, $\varphi(r=0) = \varphi_c$ and $\varphi'(r=0) = 0$ to ensure regularity at the center and to ensure the scalar field is continuous and differentiable at the surface. 

\begin{figure*}[t]
	\centering
	\includegraphics[width=7in]{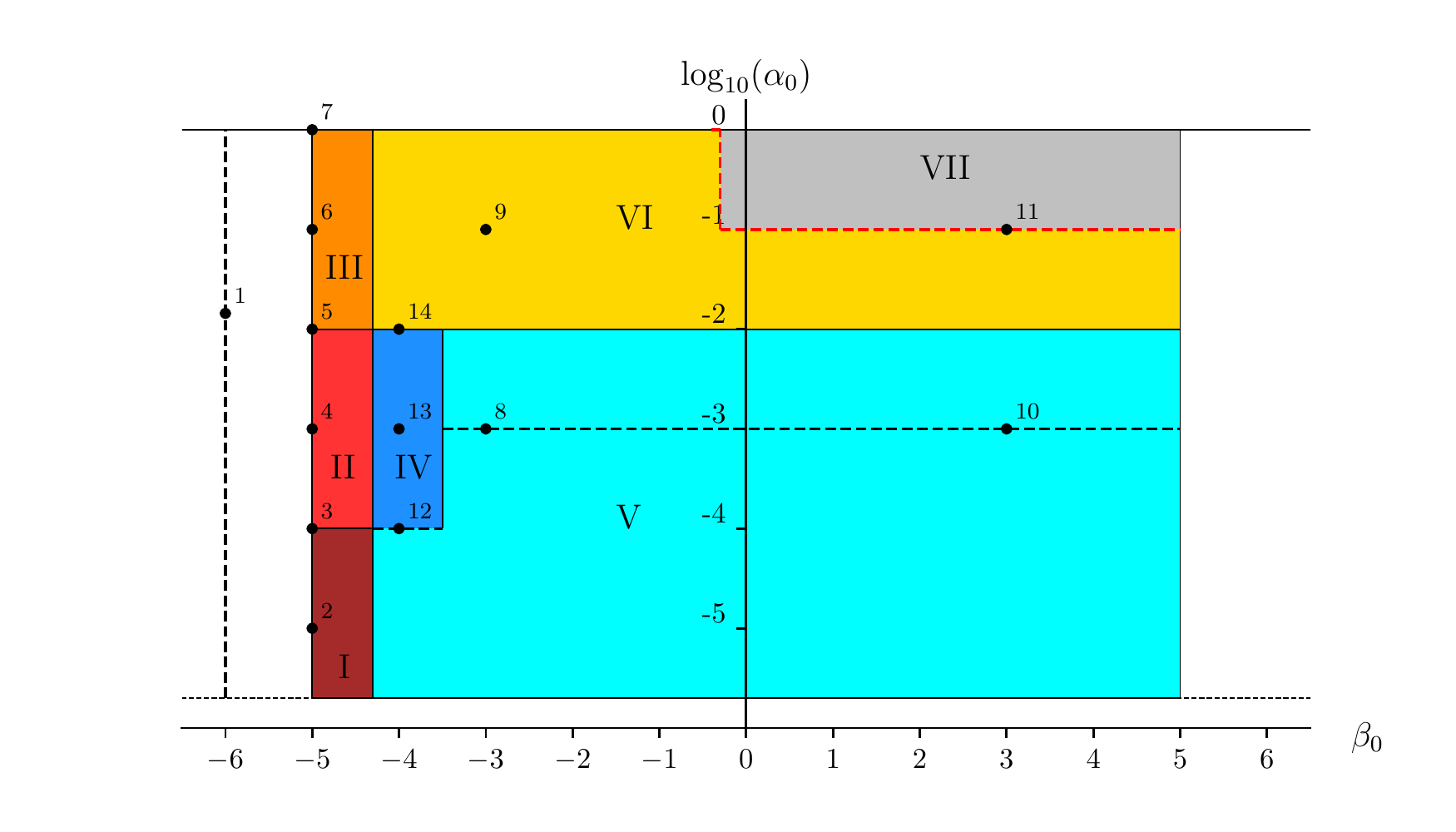}
	\caption{ \label{fig:parameter_space} A breakdown of the parameter space we explore. Each colored region, labeled also by Roman numeral, has a different numerical grid associated with it that is explain in detail throughout the text. The numbered points appearing here represent points in parameter we use to estimate the error in our numerical calculations, the details of which can be found in Appendix \ref{sec:error}.
	}
\end{figure*}

Now we must solve Eq.~(\ref{eq:analytic-DEF}) inside and outside the star, subject to the boundary conditions above. The exterior solution takes the form
\be
	\varphi \eq \varphi_\infty + \dfrac{G \omega_A}{r} + \mathcal{O}\left(\dfrac{1}{r^2}\right)\,\,,
\label{eq:weak-field-phi}
\ee
where $\varphi_\infty$ and $\omega_A$ are integration constants. There are two interesting scenarios that can occur within the star, namely when the product $\beta_0\,T_\mat$ is positive and when it is negative. In the positive case, the interior solution for the scalar field takes the form
\be
	\varphi(r<R) \eq \dfrac{\varphi_\infty}{\cos \left(CR\right)}  \dfrac{\sin\left(C\,r\right)}{C\,r}\,\,.
\label{eq:interior-phi-pos}
\ee
The asymptotic value of the scalar field at infinity $\varphi_\infty$ can made chosen to be arbitrarily small, but even in the case of $\varphi_{\infty} \rightarrow 0$, the central value of the scalar field can still be non-zero if $\cos(C R) \to 0$ at the same rate. While this is an over-simplified description of the problem, one in which we essentially are assuming a constant density inside the star and ignoring non-linear effects, it does demonstrate that there can be non-trivial scalar field solutions even when one forces the asymptotic value of the scalar field to vanish. 

In the situation where the product $\beta_0\,T_\mat$ is negative, there exists an opposite effect inside the star: any deviations from GR are exponentially suppressed. The negative case leads to an interior solutions of the form
\be
	\varphi(r<R) \eq \dfrac{\varphi_\infty}{\cosh \left(CR\right)}  \dfrac{\sinh\left(C\,r\right)}{C\,r}\,\,,
\label{eq:interior-phi-neg}
\ee
in which case any non-vanishing value of $\varphi_\infty$ is suppressed even further by $\cosh \left(CR\right)$. The suppression mechanism drives the STT solution to GR inside the star when $\beta_0\,T_\mat < 0$. 

In this linear $\varphi$ regime, the quantity $G \omega_A$ appearing in Eq.~(\ref{eq:weak-field-phi}) can be expressed as
\be
	G \omega_A \eq - \varphi_\infty \left(R - \dfrac{\tan(CR)}{C}\right)\,\,,
\label{eq:analytic-charge}
\ee
in the case of $\beta_0\,T_\mat > 0$ and with the tangent exchanged for a hyperbolic tangent when $\beta_0\,T_\mat < 0$. Recalling from Sec.~\ref{sec:background} that $\alpha_A \eq - \omega_A/m_A$ and that $\varphi_\infty = \alpha_0/\beta_0$ one finds that $\alpha_A \propto f(\beta_0, m_A) \alpha_0$\footnote{Technically, this relation should read $\alpha_A \propto f(\beta_0, m_A) \alpha_\infty$ as the scalar charge should always reduce to its weak field counter part $\alpha_\infty$ in the absence of strongly self-gravitating matter. However, these relations are already derived under weak field assumptions and in both theories $\alpha_\infty \eqsim \alpha_0$ in this regime.}, as long as one can neglect non-linear interactions of the scalar field. While many of the most interesting effects of STTs, like spontaneous scalarization, occur in the most non-linear regions of parameter space, this simple relations provides valuable insight into the types of solutions one would expect in other regions of parameter space.

%------------------------------------------------------------------------------------------------
\subsection{A classification of parameter space}\label{sec:NS:breakdown}
%------------------------------------------------------------------------------------------------

Let us now use the analytic insight described above to guide us in our numerical exploration of the $\{ \alpha_0,\, \beta_0 \}$ parameter space shown in Fig.~\ref{fig:parameter_space}. We will break our investigation into six distinct regions in parameter space, each of them having distinct features that need to be handled differently when solving for the scalar charges numerically. In all of these regions we must set up some numerical grid in $\alpha_0$, $\beta_0$, and $m_A$, and these grids are precisely how each of these regions differ from one another. We discuss the various regions in Fig,~\ref{fig:parameter_space} in detail below, relying on the analytic insight above and our numerical results, and in the next subsections we will present the numerical techniques used in each region and our results.

There exist three regions, I--III (brown, red, and orange respectively) in Fig.~\ref{fig:parameter_space} in which spontaneous scalarization occurs, i.e. when $\beta_0 \leq -4.3$. In all of these regions there exists a phase transition in the scalar field, with the sharpness of the transition being determined by the value of $\alpha_0$ (lower values leading to more steep transitions). Due to the varying level of steepness in the phase transitions and the numerical limitations of taking derivatives, we have broken this region of $\beta_0$ into 3 distinct subregions in which we use different numerical techniques to most efficiently explore the parameter space. In short, region I in Fig.~\ref{fig:parameter_space} contains the sharpest transitions, and therefore requires a finer numerical grid in $m_A$ to resolve the relevant features of interest. Region II contains less sharp transitions and it can be accurately explored with less grid points. Regions I and II both contain the same grid spacing in $\log_{10} (\alpha_0)$ and $\beta_0$, but a different spacing in $m_A$. Lastly, region III contains the same grid spacing in $m_A$ and $\beta_0$ as region II, but it contains more grid points in $\alpha_0$ to allow us to accurately calculate the numerical derivatives necessary for the scalar charges.

Region V (cyan) in Fig.~\ref{fig:parameter_space} is where the non-linearity of the scalar field can effectively be neglected and hence the scalar charge turns out to scale as in Eq.~(\ref{eq:analytic-charge}). When we solve the full set of field equations we make no approximations, but the resulting scalar charges do indeed follow the scaling relation $\alpha_A \propto f(\beta_0, m_A) \alpha_0$. Thus, in region V we solve for a single set of solutions at a single value of $\alpha_0$, lying on the black dashed lines in Fig.~\ref{fig:parameter_space} at $\alpha_0 = 10^{-3}$ for $\beta_0 > -3.5$ and $\alpha_0 = 10^{-4}$ for $\beta_0 < -3.5$, and we use the scaling relation to populate the entire region. We numerically verify in Sec.~\ref{sec:error:analytic} that these scaling relations are indeed accurate when compared to the full numerical exploration of this region of parameter space. This scalable region does not cover the entire parameter space where spontaneous scalarization does not occur. We have found numerically that the scaling does not hold when $\beta_0 < -3.5$ and when $\alpha_0 \in  (10^{-4},10^{-2})$, which is why we have isolated this part of parameter space to region IV. In this region, we investigate the solutions as if we expected spontaneous scalarization. 

For region VI in Fig.~\ref{fig:parameter_space} we find that the scaling relations previously discussed no longer exist because of how large $\alpha_0$ can become. The lack of quasi-analytic solutions here should not come as a surprise because this region has such large values of $\alpha_0$ that STT modifications can be easily constrained with solar system observations. For completeness, however, we still investigate this region extensively as some parts of this parameter space are actually useful when placing binary pulsar constraints\footnote{In Ref.~\cite{Freire:2012mg}, for example, there exists a region near $\beta_0 \sim -2$ in which constraints on scalar dipole radiation fail to constrain STTs better than Cassini and other weak field tests. This ``horn'' appearing in the binary pulsar constraints occurs for NS-WD binaries in which the quantity $(\alpha_{NS} - \alpha_{WD})^2 \sim (\alpha_{NS} - \alpha_{0})^2$ vanishes, which tends to happen near $\beta_0 \sim -2$.}. We have also removed a portion, region VII in gray, from the parameter space because we are not able to calculate NS solutions here. This has been noted in the literature before when considering DEF theory~\cite{Mendes:2016fby, Anderson:2017phb}, but in those papers, the authors only investigate small values of $\alpha_0$ and focused on values of $\beta_0$ considerably larger than what we consider here. Nonetheless, we find results very similar in this gray region of parameter space for both DEF and MO theory and it becomes impossible to extract any useful information from our numerical calculations. Therefore, since binary pulsar typically do not probe this region and Solar System tests have already ruled it out, we are justified in neglecting it.

The final point to discuss regarding the parameter space is the vertical dashed line and numbered points appearing in Fig.~\ref{fig:parameter_space}. The black vertical dashed line at $\beta_0 =-6$ marks a set of special solutions we have calculated to compare our results to those original found in Ref.~\cite{Damour:1996ke}. While we do not explore this region of parameter space in depth it provides a useful comparison to validate our code and make comparisons between known results in DEF theory and new results in MO. Details about the numbered points in Fig.~\ref{fig:parameter_space} can be found in Table~\ref{tab:special_points} and they represent a set of points we use to investigate the error in our numerical results. The details of this error analysis can be found in Sec.~\ref{sec:error}.

%------------------------------------------------------------------------------------------------

%------------------------------------------------------------------------------------------------
\subsection{Numerical methodology}\label{}
%------------------------------------------------------------------------------------------------

\begin{table*}[t!]
	\centering
	\begin{tabular*}{5.7 in}{R{.6 in} |C{.5 in} |C{.5 in} |C{.5 in} |C{.5 in} |C{.5 in} |C{.5 in} |C{.5 in} |C{.5 in} |C{.5 in} L{1 pt}}
		\hline
		\hline
		Method& -4 & -3 & -2 & -1 & 0 & 1 & 2 & 3 & 4 & \\ [1 pt]
		\hline
		central& --- & --- & 1/12 & -2/3 & 0  & 2/3 & -1/12 & --- & --- &\\ [1 pt]
		\hline
		forward& --- & --- & --- & --- & -25/4  & 4 & -3 & 4/3 & -1/4 &\\ [1 pt]
		\hline
		backward& 1/4 & -4/3 & 3 & -4 & 25/4   &---  & --- & --- & --- &\\ [1 pt]
		\hline
	\end{tabular*}
	\caption{\label{tab:coefficents} The finite difference coefficients found in Eqs.~(\ref{eq:fd_central})-(\ref{eq:fd_backward}). For every method (central, forward, or backward) the columns denote the value of the coefficient for the corresponding subscripts found in Eqs.~(\ref{eq:fd_central})-(\ref{eq:fd_backward}) .
	}
\end{table*}

We parameterize our NS solutions by a choice of $\{\jrho_c,\,\alpha_0,\,\beta_0 \}$ which ultimately determines the star's gravitational mass $m_A$, baryonic mass $\bar{m}_A$, and the asymptotic value of the scalar field $\varphi_\infty = \alpha_0 / \beta_0$. We take the approach of solving Eqs.~(\ref{eq:main-equation}) starting from the center of the NS, using Eqs.~(\ref{eq:central-conditions}) to start our numerical integration away from the singularity at $\rho=\rho_{\min}$. We follow the methods employed in Refs.~\cite{Anderson:2016aoi, Anderson:2017phb} and use {\sc{\small Mathematica}}'s default ODE solver\footnote{The default method used is LSODA which is a variant of the original LSODE (Livermore Solver for Ordinary Differential Equations) approach to solving a wide class of differential equations.} to integrate the equations to the surface of the NS where we then use Eqs.~(\ref{eq:surface-boundary}) to extract values at spatial infinity. 

In order to begin the integrations, however, we must make a choice of $\jrho_c$ and $\varphi_{c}$ as these are the two independent parameters appearing in the boundary conditions. It is near impossible to guess the correct value of $\varphi_c$ that correspond to a given $\varphi_\infty = \alpha_0/\beta_0$ to within a small tolerance. Therefore, we use a Newton-Raphson shooting method to converge onto the correct value of $\varphi_c$. In particular, we solve the equations again with a new central value of the scalar field $\varphi_{c,n+1} = \varphi_{c,n} + \Delta \varphi_c$, which gives a slightly different value of $\varphi_\infty$. The difference of the extracted values of $\varphi_\infty$ allow us to construct a simple difference equation
\be
	\varphi_{c,n+1} \eq \varphi_{c,n} - \Delta\varphi_c \dfrac{\varphi_{\infty,n} - \alpha_0/\beta_0}{\varphi_{\infty,n+1} - \varphi_{\infty,n}}\,\,,
\ee
where $n$ is the iteration number. The equation above allows us to predict a new value of $\varphi_c$ that gives a value of $\varphi_\infty$ that is closer to the desired result, and we iterate this process until the resulting value of $\varphi_\infty$ is equivalent to $\alpha_0/\beta_0$ to within some numerical tolerance. At the subsequent point in $\jrho_c$ we use the previous value of $\varphi_c$ as the starting point for the shooting, which typically allows convergence to within numerical tolerance in about 2-3 iterations.

The method described above provides NS solutions corresponding to a single combination of $\{\jrho_c,\,\alpha_0,\,\beta_0 \}$ for any choice of theory and EOS. It is convenient that the scalar charge $\alpha_A$ can be extracted directly from the boundary conditions at the surface. The quantities $\beta_A$ and $k_A$, however, must be calculated by taking derivatives across multiple solutions while keeping the baryonic mass constant. The most accurate way to take these derivatives would involve parameterizing the NS solutions by $\{\bar{m}_A,\,\alpha_0,\,\beta_0 \}$ instead and shooting in \emph{both} $\jrho_c$ and $\varphi_c$. Such an approach would allow one to construct multiple NS solutions with identical values of $\bar{m}_A$ and varying values of $\varphi_\infty$, corresponding to different values of $\alpha_0$, which is required for the derivatives needed to calculate $\beta_A$ and $k_A$. While this approach works extremely well, it is very computationally expensive since one must now shoot in two dimensions multiple times just to calculate the scalar charges for a single combination of $\{\bar{m}_A,\,\alpha_0,\,\beta_0 \}$. To completely populate the parameter space of interest one must compute the charges for roughly $10^5$ combinations of $\{\bar{m}_A,\,\alpha_0,\,\beta_0 \}$ just for a single EOS and theory choice. This quickly becomes cumbersome from a computational standpoint so we decided to take a different approach.

\begin{figure*}[t]
	\centering
	\includegraphics[width=6in]{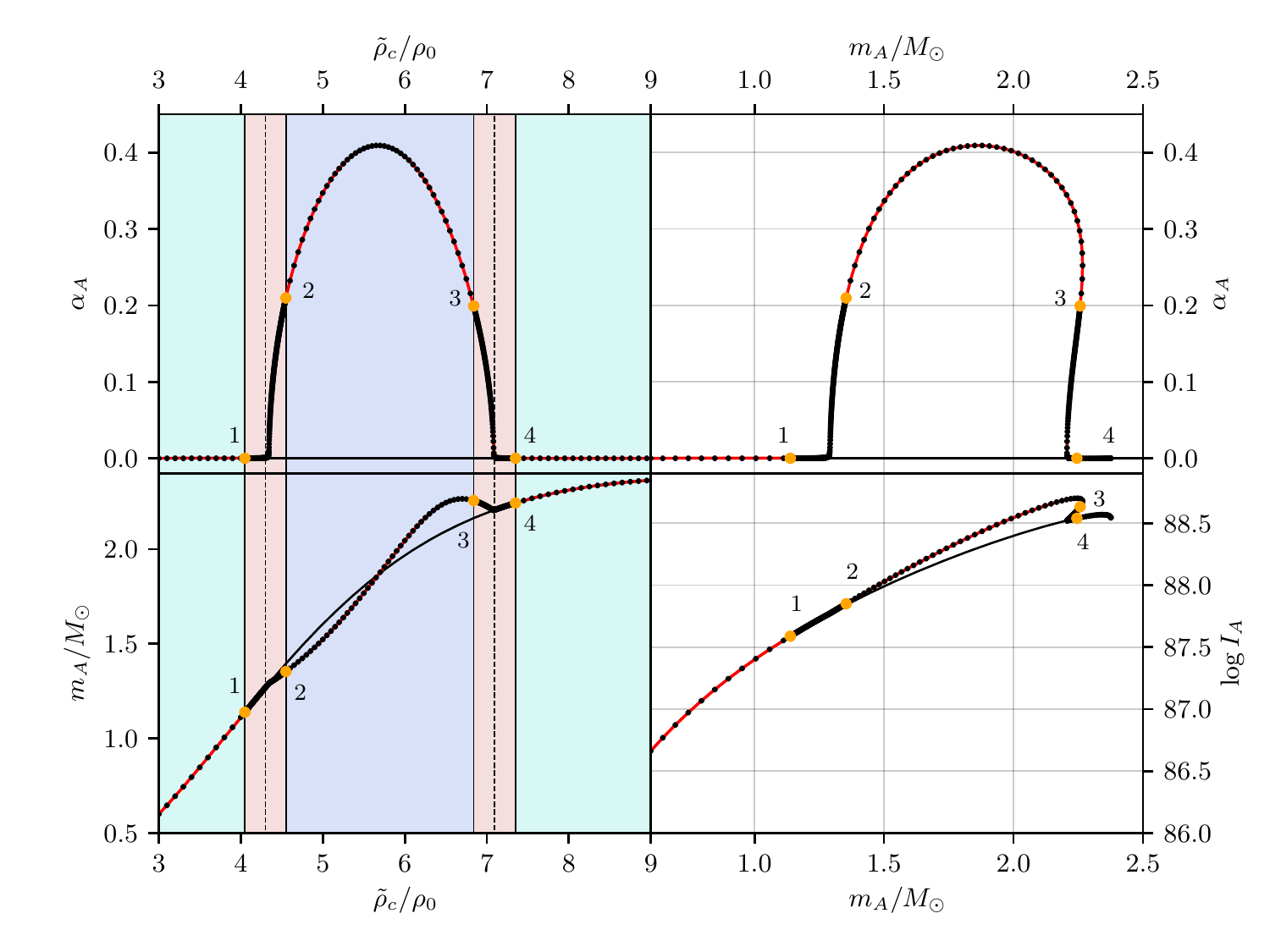}
	\caption[Numerical grid example for $m_A$]{ 
		\label{fig:mass_grid} As example of the numerical grid we use and the resulting solutions for $m_A$ for DEF theory and AP3 EOS with $\beta_0 = -5.0$ and $\alpha_0 = 10^{-5}$. Starting counter-clockwise from the top-left: $\alpha_A (\jrho_c)$, $\alpha_A(m_A)$, $\log_{10} I_A(m_A)$, and $m_A(\jrho_c)$. The orange points marked 1--4 represent the same NS solution on each of the panels and the solid black curves appearing in each panel represents the GR solution. The green regions represent the lowest resolution, followed by the blue region, and with th red regions being most dense, as described in detail in the text. The vertical dashed lines represent the critical values of central density at which point spontaneously scalarization ``turns on'' and ``turns off''.
	}
\end{figure*}

Our computational method is as follows. We continue to parameterize the NS solutions with $\{\jrho_c,\,\alpha_0,\,\beta_0 \}$, but rather than focusing on a single value of $\bar{m}_A$, we calculate an entire mass-radius (MR) curve of solutions, corresponding to a set of $\{\jrho_{c,i}\}$, for a large discrete set of $\{\alpha_0,\,\beta_0 \}$ values. This approach allows us to then interpolate the  scalar charge $\alpha_A$ as a function of the baryonic mass, thus generating curves like those in Fig.~\ref{fig:mass_grid}. Because we have a finely discretized grid in $\bar{m}_A$, we can interpolate between points and extract $\alpha_A(\bar{m}_A)$ for \emph{any} value of baryonic mass, and we can do the same for every curve we calculate. Therefore, we can compute numerical derivatives of the scalar charge (or any other quantity) at any value of baryonic mass in a computationally efficient way. While this method is prone to more numerical error than the previous one, it allows us to sample the entire parameter space very finely and it can be carried out orders of magnitude faster. A discussion of the errors associated with our methods is presented in Sec.~\ref{sec:error:analytic}

Let us now discuss the way we take numerical derivatives. We choose to use a fourth-order accurate finite difference scheme to calculate the derivatives in Eqs.~(\ref{eq:charge_2})-(\ref{eq:charge_3}). For reasons discussed below, we have to use central, forward, and backward finite difference schemes in order to most effectively utilize our numerical grid, and they take the forms
\ba
	\left(\dfrac{d F}{d \varphi_0}\right)_{c} &\eq& \dfrac{c_{\mbox{\tiny -2}} F_{\mbox{\tiny -2}} + c_{\mbox{\tiny -1}} F_{\mbox{\tiny -1}} + c_{\mbox{\tiny 0}} F_{\mbox{\tiny 0}} + c_{\mbox{\tiny +1}} F_{\mbox{\tiny +1}} + c_{\mbox{\tiny +2}} F_{\mbox{\tiny +2}}}{\Delta \varphi_0}\,\,,\nn\\ 
	\label{eq:fd_central}\\
	\left(\dfrac{d F}{d \varphi_0}\right)_{f} &\eq& \dfrac{f_{\mbox{\tiny 0}} F_{\mbox{\tiny 0}} + f_{\mbox{\tiny +1}} F_{\mbox{\tiny +1}} + f_{\mbox{\tiny +2}} F_{\mbox{\tiny +2}} + f_{\mbox{\tiny +3}} F_{\mbox{\tiny +3}} + f_{\mbox{\tiny +4}} F_{\mbox{\tiny +4}}}{\Delta \varphi_0}\,\,,\nn\\ 
	\label{eq:fd_forward}\\
	\left(\dfrac{d F}{d \varphi_0}\right)_{b} &\eq& \dfrac{b_{\mbox{\tiny -4}} F_{\mbox{\tiny -4}} + b_{\mbox{\tiny -3}} F_{\mbox{\tiny -3}} + b_{\mbox{\tiny -2}} F_{\mbox{\tiny -2}} + b_{\mbox{\tiny -1}} F_{\mbox{\tiny -1}} + b_{\mbox{\tiny 0}} F_{\mbox{\tiny 0}}}{\Delta \varphi_0}\,\,,\nn\\
	\label{eq:fd_backward}
\ea
where $F_{ \pm n} = F(\varphi_0 \pm n \Delta \varphi_0)$, and $c_i$, $f_i$, and $b_i$ are the corresponding finite difference coefficients for central, forward, and backward derivatives respectively found in Table~\ref{tab:coefficents}.

Let us now briefly discuss the numerical grid in parameter space. Each region of parameter space in Fig.~\ref{fig:parameter_space} uses a different numerical grid in $\{m_A,\, \alpha_0,\,\beta_0 \}$.  The spacing in $\alpha_0$ is determined by the level of accuracy we want when using the various finite difference schemes for the derivatives. Since $\Delta \varphi \propto \Delta \alpha_0$ we need to choose our spacing such that $\Delta \alpha_0$ between consecutive solutions branches is not too large. Finally, the $\beta_0$ grid is determined strictly by the presence of spontaneous scalarization. Therefore, if $\beta < -4.3$, the grid spacing is $\Delta \beta_0 = 0.02$ and otherwise it is $\Delta \beta_0 = 0.1$. The details of the grids in each subspace are presented in the next subsection. 

\begin{figure*}[t]
	\centering
	\includegraphics[width=7in]{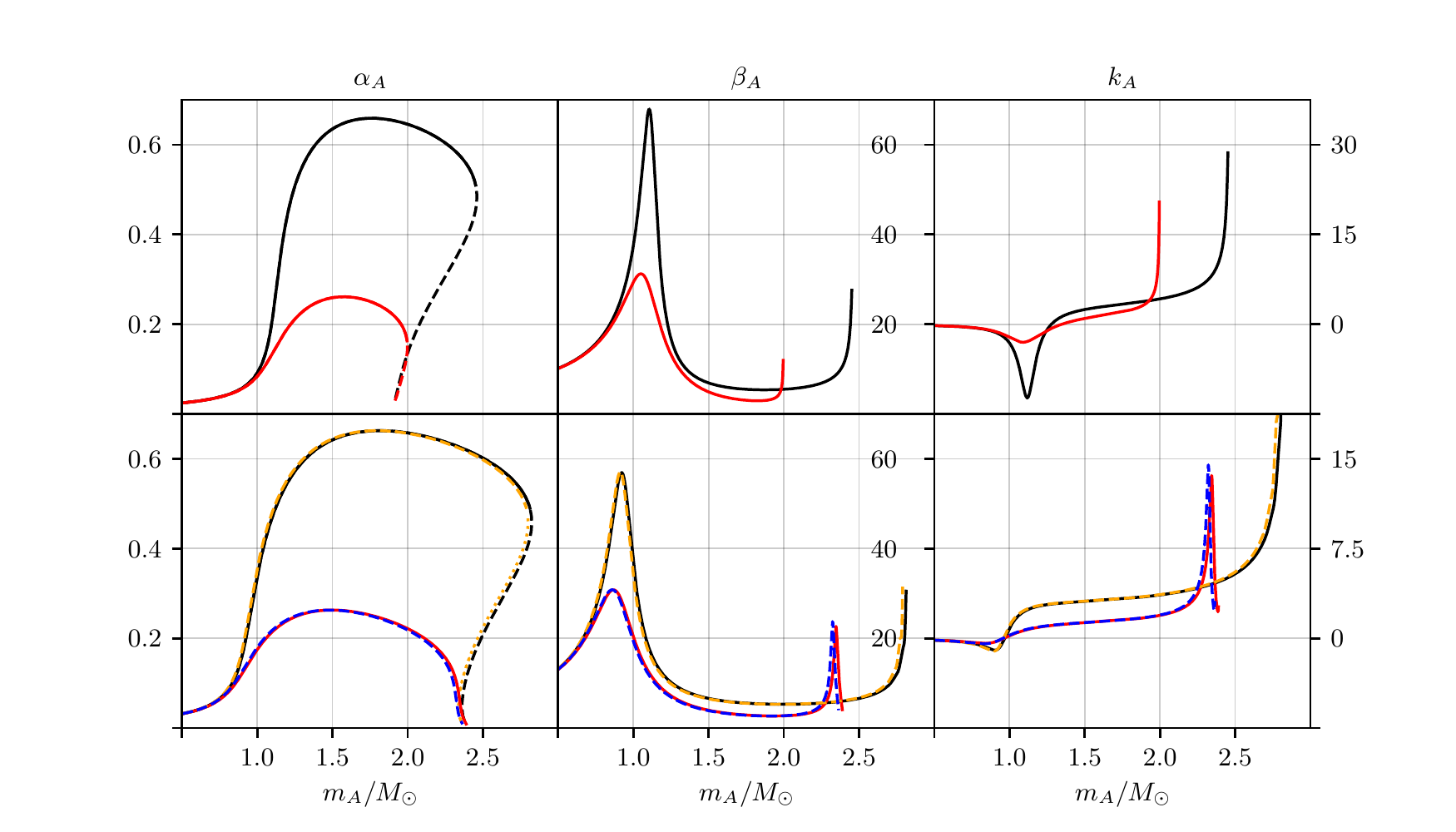}
	\caption[Numerical grid example for $m_A$]{ 
		\label{fig:def96_comp} The different scalar charges $\alpha_A$, $\beta_A$, and $k_A$, appearing in the first, second, and third column respectively, as function of the gravitational mass of the NS for $\beta_0 = -6$ and $\alpha_0 =1.44\times 10^{-2}$. The top row was obtained using the polytropic equation of state in Sec.~\ref{sec:NS:eos:polytrope}, with black curves corresponding to DEF theory and red curves to MO theory. The bottom row was obtained using the tabulated AP3 EOS (black and red) and its piecewise polytropic approximant (orange and blue) where black and orange curves are for DEF theory and red and blue for MO theory.
	}
\end{figure*}

The spacing in $m_A$ in each region is determined by the presence of any sharp features that may appear in the solutions, such as spontaneous scalarization. The green regions appearing in the left panels of Fig.~\ref{fig:mass_grid} represent the lowest resolution portions of our grid, in which we have a grid point every $0.1\rho_0$; we call this value $\Delta\rho_c^{GR}$ since it is dense enough to accurately reproduce a MR curve in GR. Spontaneous scalarization turns ``on'' and ``off'' in the red regions in Fig.~\ref{fig:mass_grid}, and in order to capture the phase transition effectively, we increase our resolution to $\Delta\rho_c^{PT} = \Delta\rho_c^{GR}/800$, where $PT$ stands for phase transition\footnote{We only increase this resolution by a factor of 800 when we consider $\alpha_0 \leq 10^{-4}$ and $\beta_0 < -4.3$. When $\alpha_0 > 10^{-4}$, we find that we do not need to sample as many points to fully capture the features of the phase transition, hence we allow $\Delta\rho_c^{PT} = \Delta\rho_c^{GR}/200$. In regions where spontaneous scalarization does not occur, i.e. $\beta > -4.3$, there is no phase transition and we simply use $\Delta\rho_c = \Delta\rho_c^{GR}/20$ everywhere.}. The blue region in Fig.~\ref{fig:mass_grid} between the phase transitions is where there are no sharp features, but where we still want increased resolution since non-linear effects do come into play; in these regions, we use a grid spacing $\Delta\rho_c^{scal} = \Delta\rho_c^{GR}/20$ where $scal$ stands for scalarization. Figure~\ref{fig:mass_grid} illustrates the density of grid points in these different regions by the number of orange circles appearing along each curve. We limit our solutions to values of $\jrho_c$ that give a $0.5 M_\odot$ NS in GR and the $\jrho_c$ that predicts the maximum mass NS in GR, for each EOS we consider.  As in the $(\alpha_{0},\beta_{0})$ grid spacing case, the details of the $\jrho_{c}$ grid are presented in the next subsection.

%------------------------------------------------------------------------------------------------

%------------------------------------------------------------------------------------------------
\subsection{Numerical results}
\subsubsection{Code Verification: the $\beta_0 = -6.0$ case}\label{sec:charges:beta0}
%------------------------------------------------------------------------------------------------

With a basic idea of our numerical grid and numerical methods for solving NS solutions in hand, let us present a comparison between our results and the ones found in~\cite{Damour:1996ke}. In that study, the authors used the polytropic equation of state described in Sec.~~\ref{sec:NS:eos:polytrope} and they show results for the $\beta_0 = -6$ and $|\varphi_\infty| = 2.4 \times 10^{-3}$ (or $\alpha_0 = 1.44\times 10^{-2}$ in our framework) case, which corresponds to point 1 in Fig.~\ref{fig:parameter_space}. In addition to using the polytropic EOS of~\cite{Damour:1996ke}, we will also show here results for the tabulated and piecewise polytropic versions of AP3 for both the DEF and MO STTs.

Let us first take a look at the top row of Fig.~\ref{fig:def96_comp}. Comparing our result to those in \cite{Damour:1996ke}, we see that the black curves, those for DEF theory, are in great agreement\footnote{Noticed that we have used gravitational mass instead of baryonic mass on the horizontal axis.}. The red curves in Fig.~\ref{fig:def96_comp} are the results for the MO theory and we see that in every case the magnitude of the charge is less than those of DEF theory and that they also do not reach very high masses. This feature is general for MO theory, i.e.~this feature is not a result of a special choice of $\alpha_0$ and $\beta_0$. The reasoning behind this is that the curvature of the conformal potential appearing in Eq.~(\ref{eq:COSH-potential}) is not as large as that of DEF theory, and therefore, the excitation of the scalar field is never as strong once the instability occurs. The curves for $\beta_A$ and $k_A$ formally diverge at higher masses, a phenomenon due to the fact that these are derivatives of quantities that ``turn over'' on themselves, e.g. in the right panels Fig.~\ref{fig:mass_grid} one can see that the slope in $\alpha_A$ and $\log_{10} I_A$ become infinite at some point as the mass increases.

\begin{figure*}[t]
	\centering
	\includegraphics[width=7in]{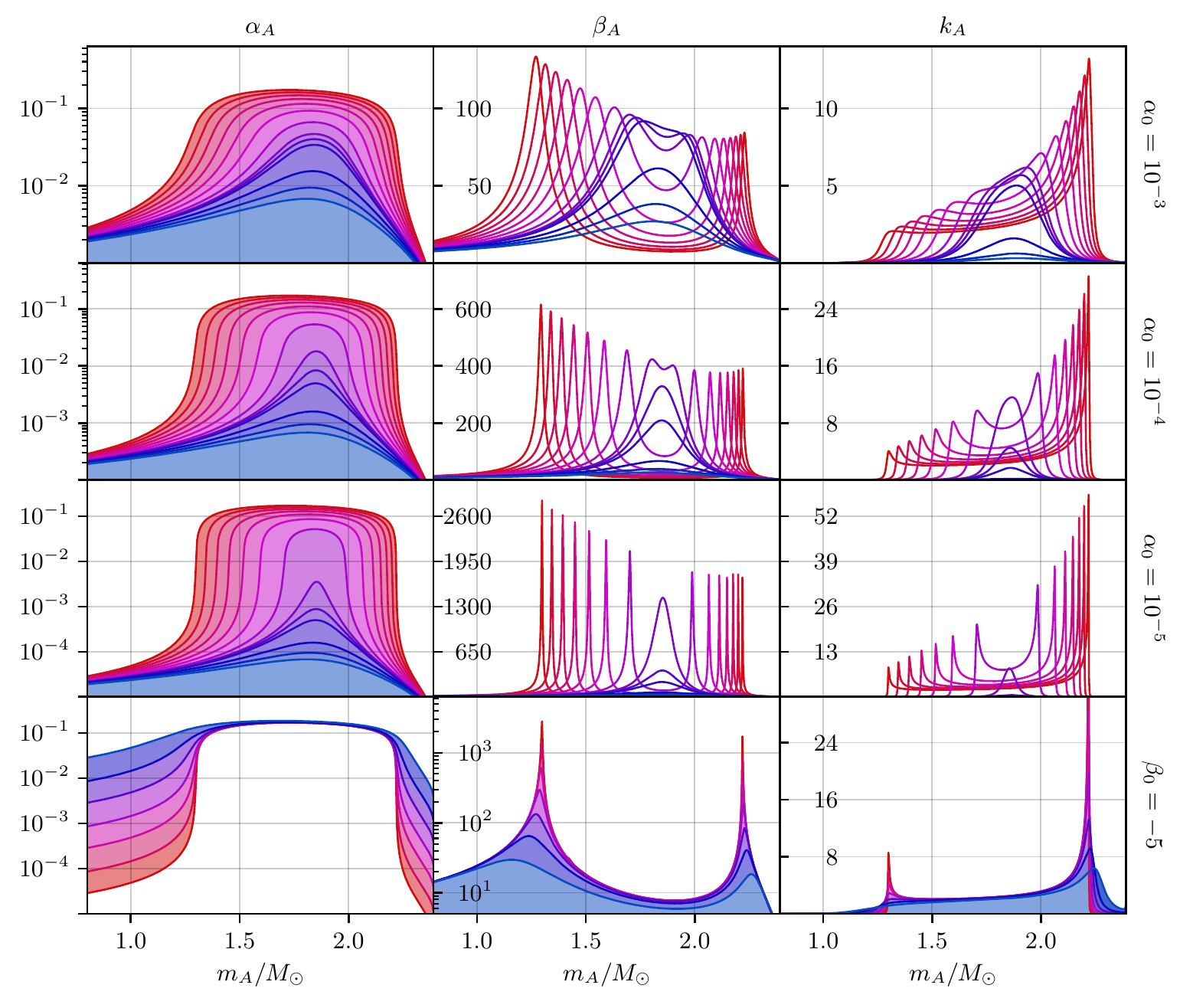}
	\caption[Numerical grid example for $m_A$]{ 
		\label{fig:ss} The behavior of the scalar charges as one changes $\alpha_0$ and $\beta_0$. Here we have used MO theory with AP3 EOS. The first three rows correspond to constant values of $\alpha_0$ (show on the right) for $ -5.0 \leq \beta_0 \leq -4.0$ , ranging in color from most red to most blue respectively. The bottom row corresponds to $\beta_0 = -5$ and $ -5 \leq \log_{10} \alpha_0 \leq -2$, with colors from red to blue respectively. The shaded regions represent the smooth transitions in the solutions from one curve to the next and they continue to follow these trends as one continues to change the parameters $\alpha_0$ and $\beta_0$.
	}
\end{figure*}

The bottom row in Fig.~\ref{fig:def96_comp} shows the same scalar charges but for realistic equations of state, namely AP3 here, and its piecewise polytropic approximate. The maximum values of $\alpha_A$ are very weakly affected by the EOS in both theories. There is, however, a strong dependence on the EOS when it comes to the location of the critical mass, $m_\crit$, at which spontaneous scalarization occurs, and the maximum mass above which stable NSs do not exist. The other scalar charges, $\beta_A$ and $k_A$, are quite different between the different EOSs and theories. The phase transition is now less sharp, and therefore, the magnitude of $\beta_A$ is smaller than in the polytropic case. Moreover, it happens to be the case here that in MO theory there is no formal divergence in $\beta_A$ at large masses. Finally, the ``negative spike'' in $k_A$ that usually occurs is greatly suppressed for realistic EOSs. 

The piecewise polytropic EOS leads to scalar charges that are very similar to those found with a tabulated EOS. Aside from a very slight shift in the masses, the structure and magnitude of the scalar charges are nearly identical. The small differences are likely due to the fact that polytropes are just too simple to accurately capture the different physics that occurs at different densities inside NSs, which can over/under exaggerate features we find in the scalar charges. Using the piecewise polytropes, however, can speed up numerical calculations immensely since they are analytic.  A more detailed investigation of the differences between tabulated and piecewise polytrope results can be found in Sec.~\ref{sec:error:peicewise}

%------------------------------------------------------------------------------------------------

%------------------------------------------------------------------------------------------------
\subsubsection{Spontaneous scalarization: $\beta_0 \leq -4.3$}\label{sec:charges:beta1}
%------------------------------------------------------------------------------------------------

As is well-known in the literature, spontaneous scalarization occurs in STTs when $\beta<-4.3$ regardless of the EOS. As demonstrated in the previous section, there exists a $m_\crit$ at which a phase transition of the scalar field occurs, but the precise value of this mass, however, depends on the theory, the EOS, and the value of $\beta_0$. These dependencies require that we determine what this critical mass is, and to place a finer grid in $\jrho_c$ centered near this location to ensure we capture the details of the phase transition with enough accuracy and precision to calculate the scalar charges at these points, cf. Fig.~\ref{fig:mass_grid}. Spontaneous scalarization occurs in regions I--III (brown, red, and orange) of Fig.~\ref{fig:parameter_space}, each of which has a slightly different grid that we discuss in detail next.

%\subsubsection{Brown Region}\label{sec:charges:beta1:brown}
\vspace{0.4cm}
\textbf{Region I (Brown):}
\vspace{0.2cm}

This region has the finest grid in $\jrho_c$ because the phase transition is most sharp here, and near the transition we decrease our spacing in $\jrho_c$ by a factor of 800 relative to $\Delta\rho_c^{GR}$, as mentioned earlier. This level of resolution requires around 600 NS solutions in total for each value of $\alpha_0$ and $\beta_0$. In this region, we use $\alpha \in \{ 1.2,\, 1.1,\, \cdots ,\,0.2 \} \times 10^{\text{mag}}$ where $\text{mag} \in \{-4,\,-5\}$, and a spacing in $\beta_0$ of $\Delta \beta_0 = 0.02$. Spacing $\alpha_0$ in this manner gives a spacing in $\varphi$ of $\Delta \varphi = 10^{\text{mag}-1} /\beta_0 \sim 2\times 10^{\text{mag}}$ for this range of $\beta_0$. Since our finite difference schemes are all fourth-order accurate, our spacing in $\alpha_0$ is small enough to confidently calculate the needed derivatives\footnote{Going to smaller values of $\alpha_0$ is difficult because we can never let it change signs when calculating the derivatives. This means that $\Delta\alpha_0$ \emph{must} always be smaller than $\alpha_0$, and it is computationally expensive to calculate solutions with enough accuracy and precision for $\Delta\alpha_0 \lesssim 10^{-6}$.}.
This particular spacing allows us to use the central finite difference scheme for mantissa of $\alpha_0$ from 1.0 to 0.4, the forward finite difference scheme for 0.2 and 0.3, and the backward scheme for 1.2 and 1.1. Since we need to calculate 1.2, 1.1, 0.3, and 0.2 for using the central finite difference scheme, we automatically get the derivatives at these extra points for free just by changing the finite differencing. 

%\subsubsection{Red Region}\label{sec:charges:beta1:red}
\vspace{0.4cm}
\textbf{Region II (Red):}
\vspace{0.2cm}

In this region, we are able to use only a factor of 200 more points near the transition regions, i.e. $\Delta\rho_c^{GR} / 200$. We use the same grid in $\beta_0$ here as in region I above, i.e. $\Delta \beta_0 = 0.02$. Our grid in $\alpha_0$ is set up in a similar way as well, i.e. $\alpha \in \{ 1.2,\, 1.1,\, \cdots ,\,0.2 \} \times 10^{\text{mag}}$ where $\text{mag} \in \{-2,\,-3\}$, and are both set up to make efficient use of the finite differences introduced above. As one can see in Fig.~\ref{fig:def96_comp} for example, the phase transitions are not as sharp in this region as they are in region I (see Fig.~\ref{fig:mass_grid}), and this allows us to confidently under-sample $m_A$ relative to these smaller values of $\alpha_0$.

%\subsubsection{Orange Region}\label{sec:charges:beta1:orange}
\vspace{0.4cm}
\textbf{Region III (Orange):}
\vspace{0.2cm}

In this region, the spacing in $\alpha_0$ is finer than what we used before. Overall, we adopt a similar scheme as before, but we now require a finer grid centered around the main values of $\alpha_0$ we are interested in. Here we use $\alpha_0 \in \{ 1.0,\, 0.9,\, \cdots ,\,0.2 \} \times 10^{\text{mag}}$ with $\text{mag} \in \{0,\,-1\}$ as the main grid points, and around each of these grid points we choose the set $\alpha_{0} \in \{+2,+1,-1,-2\}\,\times 10^{\text{mag}-3}$ to give us the other necessary points needed for the finite differences. These choices roughly enforce the same level of accuracy in our results when compared to the grids used for smaller values of $\alpha_0$.

%\subsubsection{Blue Region}\label{sec:charges:beta2}
\vspace{0.4cm}
\textbf{Region IV (Blue):}
\vspace{0.2cm}

This region of Fig.~\ref{fig:parameter_space} extends to larger values of $\beta_0$ than where spontaneous scalarization occurs, but there are still non-linear effects here that come into play and prevent us from using the quasi-analytic relations. In this region, we only sample in intervals of $\Delta\beta_0 = 0.1$, but we must sample $\alpha_0$ on a grid like that used in region I. Thus, region IV is essentially a transition region between spontaneous scalarization and the rest of parameter space, and thus it requires special consideration. 

%\subsubsection{Results}\label{sec:charges:beta1:results}
\vspace{0.4cm}
\textbf{Scalar Charges:}
\vspace{0.2cm}

Some representative results for the various scalar charges in MO theory can be found in Fig.~\ref{fig:ss}, for multiple values of $\alpha_0$ and $\beta_0$ described in the caption. The first three rows in Fig.~\ref{fig:ss} show the behavior of the scalar charges for $-5 \leq \beta_0 \leq -4.0$, ranging from most red to most blue respectively, and three orders of magnitude in $\alpha_0$. Notice that as $\alpha_0$ decreases, the growth of $\alpha_A$ becomes more rapid as the phase transition of the scalar field becomes more sudden. As a result of this, the peaks in $\beta_A$ and $k_A$ increase in magnitude and decrease in width because higher order effects become more localized in $m_A$. One also notices that the location of $m_\crit$, the mass at which spontaneous scalarization turns ``on'', moves towards larger masses as $\beta_0$ become less negative, as one would expect~\cite{Damour:1996ke}. 

The maximum value of $\beta_A$ at $m_\crit$ decreases as we decrease $|\beta_0|$, while the max values of the $k_A$ tends to increase. A possible explanation for this is related to the location of $m_\crit$. For the situations where $m_\crit$ is larger, the NSs at this mass have a larger moment of inertia, and therefore, it is possible the NS's inertia is more sensitive to the external scalar field, and hence the increase in $k_A$. This reasoning also explains why the right peaks of $k_A$ are larger than the left peaks\footnote{The appearance of this second peaks is somewhat unique to MO theory for this range of $\beta_0$. As one can see in Fig.~\ref{fig:def96_comp}, $\beta_A$ and $k_A$ diverge for high masses in DEF theory but not in MO theory. For more negative values of $\beta_0$, however, MO theory also exhibits such divergences for large masses.}. 

We have determined numerically that the max values $\beta_{A,\max}$ and $k_{A,\max}$ follow simple linear relations in $\log_{10}$ space for $\alpha_0$ and a \emph{fixed} $\beta_0$. The coefficient of these relations remain EOS and theory dependent, but they take the general form
\ba
	\log_{10}\beta_{A,\max} \eq B_0(\beta_0) + B_1(\beta_0) \log_{10}\alpha_0\,\,,\\
	\log_{10}k_{A,\max} \eq  K_0(\beta_0) + K_1(\beta_0) \log_{10}\alpha_0\,\,,
\label{eq:charges23_max}
\ea
where $B_0,\,B_1,\,K_0,$ and $K_1$ are coefficients that depend on the value of $\beta_0$. The quantities $B_1$ and $K_1$ are always negative for \emph{all} values of $\beta_0$ that we consider, and there is no reason to think that this would be any different for more negative values of $\beta_0$. This means that $\beta_{A,\max}$ and $k_{A,\max}$ \emph{always} increase (decrease) with decreasing (increasing) $\alpha_0$ as one might expect. Moreover, for our data we find that $B_1$ is always less than 2, and it is typically less than unity. This is important because the combination $\alpha_0^2 \beta_A$ appears directly in the PPK parameter $\dot{\omega}$ for binary pulsars with a companion white dwarf. While it is hard to numerically investigate regions of parameter space with $\alpha_0 < 10^{-5}$, these linear relations tell us that the combination $\alpha_0^2 \beta_A$ will always decrease with decreasing $\alpha_0$ and have negligible effects on the PPK parameters. These relations only hold for $\alpha_0 < 10^{-2}$ but they provide a convenient way to estimate the maximum value of the scalar charges without numerically solving the field equations.

The bottom row of Fig.~\ref{fig:ss} shows the behavior of the charges for a constant $\beta_0 = -5$, but multiple orders of magnitude in $\alpha_0$, red corresponding to the smallest values and blue to the largest. As one may expect, the overall magnitude of $\alpha_A$ is determined by the values of $\beta_0$, while its growth rate near the critical mass is determined by $\alpha_0$ and is correlated directly with the magnitude of $\beta_A$. A similar statement can be made for $k_A$ in that $\alpha_0$ determines how sudden the growth of the scalar field is, and therefore, it leads to larger values of $k_A$ near the critical mass at which these transitions occur.
%------------------------------------------------------------------------------------------------

%------------------------------------------------------------------------------------------------
\subsubsection{No Spontaneous Scalarization: $\beta_{0} > -4.3$}\label{sec:charges:beta3}
%------------------------------------------------------------------------------------------------

%\subsubsection{Cyan Region}\label{sec:charges:beta3:cyan}
\vspace{0.4cm}
\textbf{Region V (Cyan):}
\vspace{0.2cm}

This region of parameter space is perhaps the simplest and easiest to explore numerically. As mentioned in Sec.~\ref{sec:NS:analytic}, as long as $\alpha_0$ is not too large, there exist scaling relations that we can use to calculate the scalar charges. Reiterating what we explained in that section, in this region of parameter space we can assume the scalar charge $\alpha_A$ takes the form 
\be
	\alpha_A \eq \alpha_0 f(\beta_0, m_A)\,\,,
\label{eq:charge1_scale}
\ee
which then allows us to derive simple relations for the other scalar charges. Taking the derivative of this scalar charge, and applying the chain rule to the definition of $\beta_A$, we find
\be
	\beta_A \eq \dfrac{\partial \alpha_A}{\partial \alpha_0} \left(\dfrac{\partial \varphi_\infty}{\partial \alpha_0}\right)^{-1} \eq \beta_0 \, f(\beta_0, m_A) \eq \dfrac{\alpha_A}{\varphi_\infty}\,\,.
\label{eq:charge2_scale3}
\ee
Equation~(\ref{eq:charge2_scale3}) tells us that, for any value of $\beta_0$, $\beta_A (m_A)$ is the same for all values of $\alpha_0 \lesssim 10^{-2}$. Moreover, this equation also tells us that $\beta_A$ is always directly proportional to $\alpha_A$, meaning that we technically do not even need to take any derivatives to determine it. 

We can find a similar relation to that in Eq.~(\ref{eq:charge1_scale}) for the inertial charge $k_A$, but the derivation is slightly more complicated. Starting with Eqs.~\eqref{eq:BC-J}-\eqref{eq:BC-Omega} and assuming weak fields everywhere such that $e^{\hat{\nu}} \sim 1$ one finds that the moment of inertia becomes
\be
I \eq \dfrac{J}{\Omega} \approx \dfrac{G m^2 R (3 - \alpha_A^2)}{3 c^2}\,\,,
\ee
where we have also neglected terms of order $(Gm/Rc^2)^2$. Using this in the definition for $k_A$ in Eq.~\eqref{eq:charge_3} we find
\be
k_A \approx \alpha_A\left( 2 + \dfrac{m}{R}\dfrac{\partial R}{\partial m} - \dfrac{2}{3}\beta_A \right)\,\,,
\ee
where we have made use of the definitions of the other scalar charges $\alpha_A$ and $\beta_A$. Then, making use of Eqs.~\eqref{eq:charge1_scale}-\eqref{eq:charge2_scale3} we find that
\be
k_A \eq \alpha_0\,g(\beta_0, m_A)\,\,,
\label{eq:charge3_scale}
\ee
where $g$ is a function independent of $\alpha_0$.

\begin{figure}[t]
	\centering
	\includegraphics[width=3.5in]{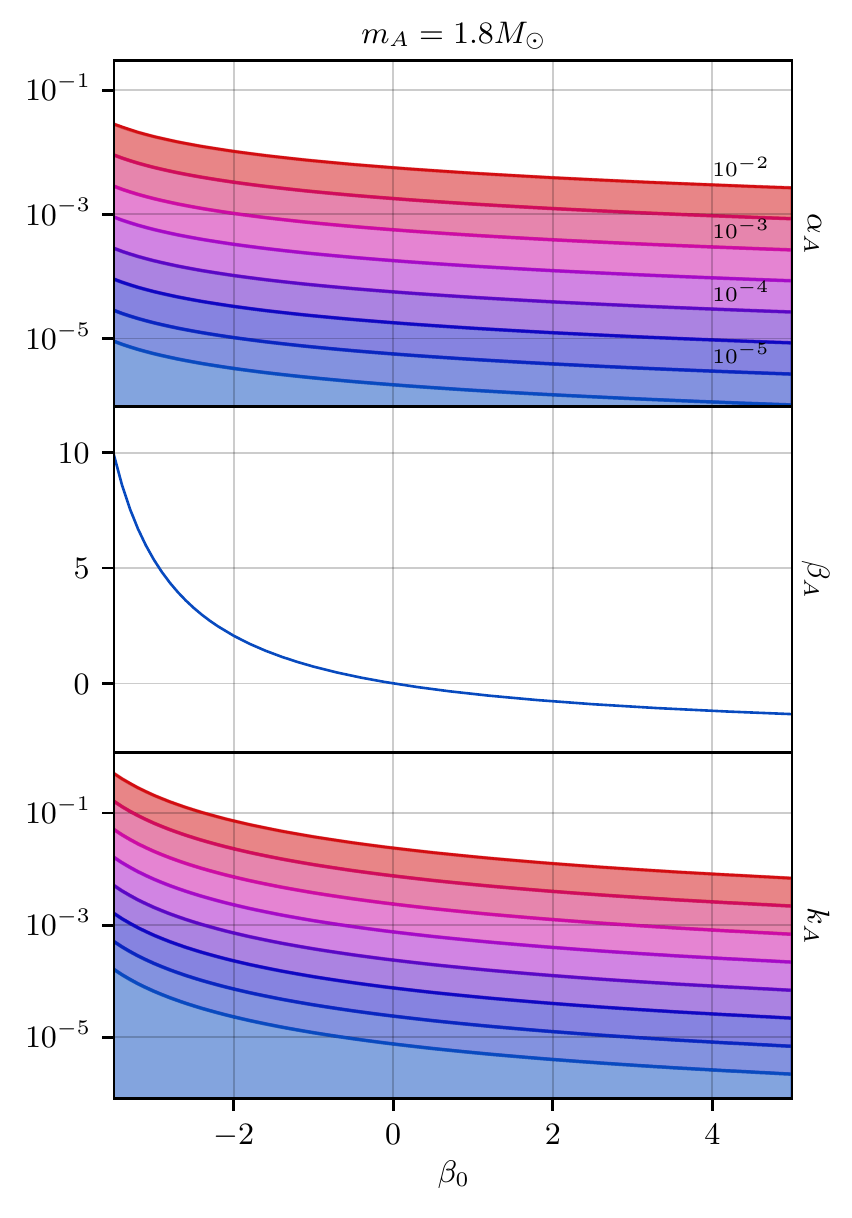}
	\caption[Numerical grid example for $m_A$]{ 
		\label{fig:error_data_file3} An example of the scaling relations described in the text for MO theory and AP3 EOS. The scalar charges $\alpha_A$ and $k_A$ do indeed scale directly with $\alpha_0$ and $\beta_A$ behaves independent of the value of $\alpha_0$.
	}
\end{figure}

The scaling relations we have introduced make the exploration of this region in $\beta_0$ almost trivial. We simply calculate the three scalar charges for points in parameter space lying on the horizontal dashed lines of Fig.~\ref{fig:parameter_space} (at $\alpha_0 = 10^{-4}$ and $\alpha_0 = 10^{-3}$), and then, we rescale the results to find the charges for any other point in region V, holding $\beta_0$ constant. For the solutions we do calculate directly, we use a grid in $\beta_0$ given by $\Delta \beta_0 = 0.01$ and a grid in $\jrho_c$ given by $\Delta\rho_c^{GR} / 20$. Because there are no sharp features in the scalar charges in this region of parameter space, we sample considerably less central densities when compared to when spontaneous scalarization occurs.

\begin{figure}[h!]
	\centering
	\includegraphics[width=3.5in]{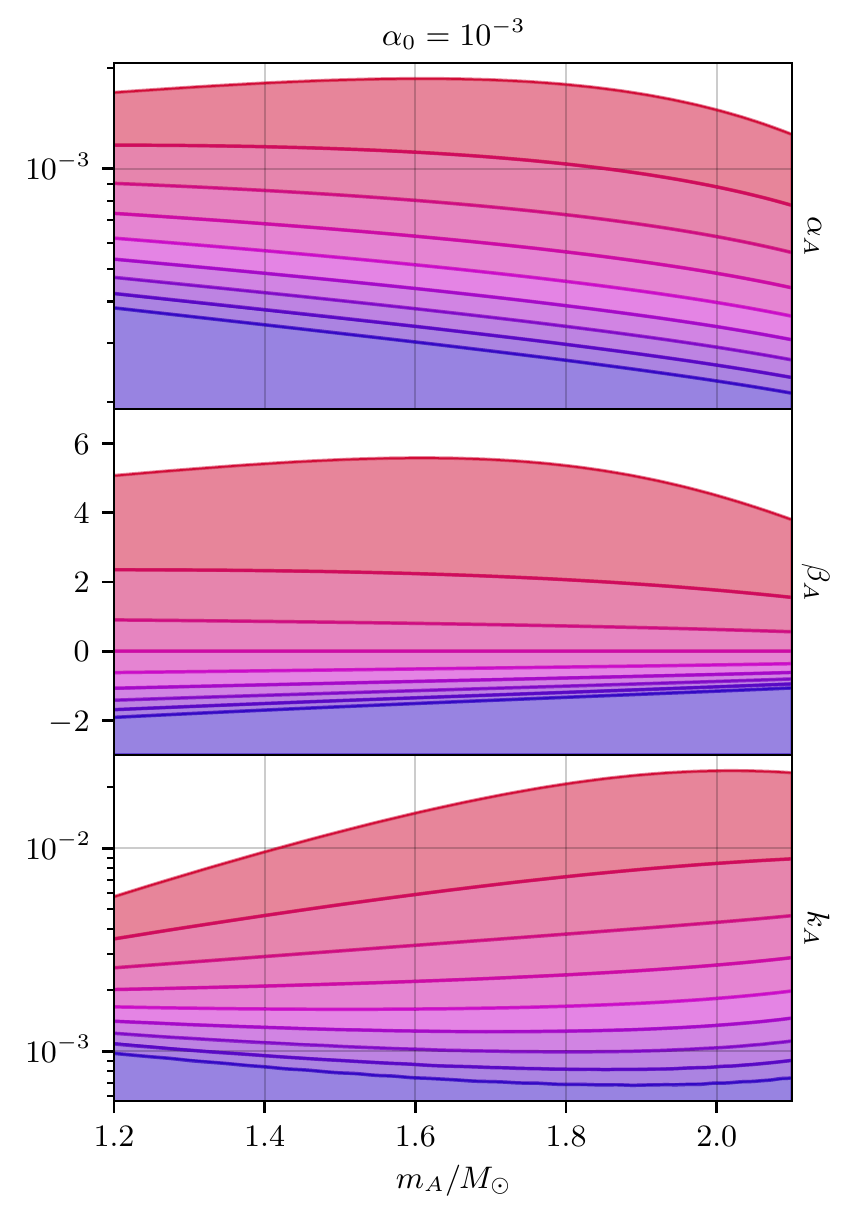}
	\caption[Numerical grid example for $m_A$]{ 
		\label{fig:error_data_file4} Examples of the scalar charges found in the cyan region of parameter space in Fig.~\ref{fig:parameter_space}, for MO theory and AP3 EOS. We include curves for $-3 \leq \beta_0 \leq 5$ with spacing $\Delta \beta_0 = 1$, with color ranging from red to blue respectively.
	}
\end{figure}

%\subsubsection{Yellow Region}\label{sec:charges:beta3:yellow}
\vspace{0.4cm}
\textbf{Region VI (Yellow):}
\vspace{0.2cm}

This region presents the same numerical difficulties as region III, expect that there is a lack of spontaneous scalarization in region VI. We use the same grids in $\beta_0$ and $\alpha_0$ as those used in region III, and the same grid in $\jrho_c$ as that used in region V. Note, however, the the gray region in Fig.~\ref{fig:parameter_space} cuts out a significant portion of region VI. As we mentioned earlier, this is because we cannot find stable NS solutions in the gray region, and we must thus omit them from our analysis since they are extremely unlikely to affect binary pulsars constraints.

%\emph{any} of the finite difference schemes introduced earlier in this section we find
%%
%%
%\be
%	\beta_A \eq \left.\dfrac{\partial \alpha_A}{\partial \varphi_\infty}\right|_{\bar{m}_A} \eq \dfrac{1}{\Delta \varphi_\infty} \sum\limits_{i}C_i\,\,\alpha_{A,i}\,\,,
%\label{eq:charge2_scale}
%\ee
%%
%%
%where $C_i$ are finite difference coefficients. Making use of Eq.~(\ref{eq:charge1_scale}) and noting that $\Delta \varphi_\infty = \Delta \alpha_0 / \beta_0 = \alpha_0 \times \text{const}$ for our numerical scheme\footnote{This dependence on $\alpha_0$ is actually independent of the way one chooses $\Delta \varphi_\infty$ as one can always make it proportional to $\alpha_0$ times some constant, but it most clearly seen in our description above.}, we find
%%
%%
%\be
%	\beta_A \eq \dfrac{\alpha_0}{\alpha_0 \times \text{const}} \sum\limits_{i}C_i\,\,f (\beta_0, m_A) \eq F(\beta_0, m_A)\,\,,
%\label{eq:charge2_scale2}
%\ee
%%
%%
%where $F$ is a function containing all information about $\beta_0$ and $m_A$ and is completely independent of $\alpha_0$. Alternatively, one can 

%------------------------------------------------------------------------------------------------

%%------------------------------------------------------------------------------------------------
%\subsection{Global properties of scalar charges}\label{sec:charges:properties}
%%------------------------------------------------------------------------------------------------
%
%
%
%

\vspace{0.4cm}
\textbf{Scalar Charges:}
\vspace{0.2cm}

Some representative results of the scalar charges in region V are shown in Fig.~\ref{fig:error_data_file3}. In this figure, we hold $m_A$ constant and plot the scalar charges as a function of $\beta_0$ spanning the entire region for $\beta_0 \geq -3.5$. Each curve in Fig.~\ref{fig:error_data_file3} represents a different value of $\alpha_0$ and it becomes clear that $\log_{10}\alpha_A$ and $\log_{10}k_A$ scale \emph{directly} with $\log_{10} \alpha_0$, showing that the relations in Eqs.~(\ref{eq:charge1_scale}) and (\ref{eq:charge3_scale}) are indeed accurate. Moreover, the value of $\beta_A$ shown in the middle panel of Fig.~\ref{fig:error_data_file4} shows no dependence on $\alpha_0$, verifying that Eq.~\ref{eq:charge2_scale3} holds. Figure~\ref{fig:error_data_file4} shows similar scalar charges but as a function of the gravitational mass of the NSs for a representative set of $\beta_0$ from region V. One will notice that NSs in this region of parameter space begin to ``de-scalarize'' as the mass of the NS increases, i.e. $\alpha_A$ becomes smaller than $\alpha_0$ as $m_A$ increases. As one might expect from Fig.~\ref{fig:error_data_file3}, all three scalar charges monotonically decrease in magnitude, for all masses, as one increases $\beta_0$.

%------------------------------------------------------------------------------------------------
\section{Using the data file}\label{sec:data}
%------------------------------------------------------------------------------------------------

Now that we have discussed how we calculate the charges and presented some of the results, let us discuss how one can use the end product of this analysis: the data generated for all the scalar charges. This section explains how the master data file is generated, what its properties and limitations are and

%--------------------------------------------------------------------------------
\subsection{The generation of the master data file}

We first set up our numerical grid according to Sec.~\ref{sec:charges} and subsections therein. From each NS solution we extract the boundary conditions in Eqs.~(\ref{eq:surface-boundary}) and save them to file for post-processing. The previous step requires the bulk of the computational time, as we need to calculate on the order of $10^5$ different NS solutions (1 for each combination of $\{ \jrho_c,\,\alpha_0,\,\beta_0 \}$) for \emph{each} combination of theory (DEF or MO) and EOS in Sec.~\ref{sec:NS:eos:tabulated}.

With the full set of data in hand for a theory-EOS combination, we now process it to extract the scalar charges. As we mentioned in Sec.~\ref{sec:charges}, we interpolate the raw data in order to extract information from more masses than we actually sampled. To do this interpolation, however, we need to proceed with caution when dealing with NS that spontaneous scalarize, like those in Fig.~\ref{fig:def96_comp}. One notices that $\alpha_A$ ``turns over'' on itself for large masses. While this feature is not present for every set of NSs that undergoes spontaneous scalarization, it does present a problem for interpolation since the function is multi-valued. To avoid this issue, we remove the data points that lie on the unstable branch of solutions, i.e. the ones that coincidentally make $\alpha_A$ double valued (cf. the dashed points in the top left panel of Fig.~\ref{fig:def96_comp}). This is possible because, at least when spontaneous scalarization occurs, there is \emph{always} NS solutions that reach maximum masses that are \emph{at least} as large as the maximum mass in GR\footnote{This can be seen from a mass-density curve like that in Fig~\ref{fig:mass_grid}, in which case the scalarized branch of solutions ``departs'' from the GR curve and eventually ``return'' for larger values of $\jrho_c$. Therefore, even if the scalarized branch does not produce a NS with mass greater than the maximum mass in GR, the GR branch will.}. With the unstable points of the solution removed we simply continue with the interpolation as described in the previous paragraph.

\begin{table*}[t!]
	\centering
	\renewcommand{\arraystretch}{1.3}
	\begin{tabular*}{6 in}{C{0.7 in} C{.7 in} C{.7 in} C{.7 in} C{.7 in} C{.7 in} C{.7 in} C{.7 in} C{.7 in}  C{1 pt}}
		$\alpha_A$	&	$\beta_A$	&	$k_A$		&	$\bar{m}_A$	&	$m_A$	&	$\alpha_0$	&	$\log_{10} \alpha_0$	&	$\beta_0$	&	 \\ [1 pt] 
		\hline
		\hline
		0.444979	&	0.270288	&	0.931259	&	1.16903	&	1.000	&	1.	&	0.	&	-5.00	&	 \\ [2 pt] 
		0.444787	&	0.270318	&	0.931345	&	1.17155	&	1.002	&	1.	&	0.	&	-5.00	&	 \\ [2 pt]
		0.444595	&	0.270350	&	0.931432	&	1.17408	&	1.004	&	1.	&	0.	&	-5.00	&	 \\ [-1 pt]
		\vdots		&	\vdots		&	\vdots		&	\vdots	&	\vdots	&	1.	&	0.	&	-5.00	&	 \\ [2 pt]
		0.331799	&	0.410289	&	1.101105	&	2.67262	&	2.100	&	1.	&	0.	&	-5.00	&	 \\ [2 pt]
		0.440000	&	0.314515	&	0.925646	&	1.15792	&	1.000	&	0.9	&	-0.045757	&	-5.00	&	 \\ [2 pt]
		0.439811	&	0.314451	&	0.925772	&	1.16041	&	1.002	&	0.9	&	-0.045757	&	-5.00	&	 \\ [-1pt]
		\vdots		&	\vdots		&	\vdots		&	\vdots	&	\vdots	&	\vdots	&	\vdots	&	-5.00	&	 \\ [2 pt]
		0.120843	&	10.909670	&	4.323983	&	2.50472	&	2.100	&	0.000002	&	-5.69897	&	-5.00	&	 \\ [2 pt]
		0.444997	&	0.270390	&	0.931196	&	1.16931	&	1.000	&	1.	&	0.	&	-4.98	&	 \\ [-1pt]
		\vdots		&	\vdots		&	\vdots		&	\vdots	&	\vdots	&	\vdots	&	\vdots	&	\vdots	&	 \\ [-1pt]
		\vdots		&	\vdots		&	\vdots		&	\vdots	&	\vdots	&	\vdots	&	\vdots	&	\vdots	&	 \\ [-1pt]
		0.0000004	&	1.062219	&	-0.000001	&	2.50430	&	2.100	&	0.000002	&	-5.69897	&	5.00	
	\end{tabular*}
	\caption{\label{tab:data_file} An example of the layout for the first 9 columns of the data file for MO theory and AP3 EOS. For a given value of $\beta_0$ we tabulated the data for every value of $\alpha_0$, of which for every value of $\alpha_0$ we tabulate data for all values of $m_A$ using the grid described in the text. This pattern repeats for all values of $-5 \leq \beta_0 \leq +5$ that we sampled. The definitions of the quantities in the columns is described in the text. The actual data files we provide also have data for the central density, radius, central scalar field, and surface scalar field of the NSs.
	}
\end{table*}

In total we must interpolate 4 separate functions to give us the data we need for constructing the data files, which are $\{ m_A(\bar{m}_A),\,\bar{m}_A(m_A),\,\alpha_A(\bar{m}_A) ,\, \log I_A(\bar{m}_A)\}$. We need $\alpha_A$ and $\log I_A$ as function of the baryonic mass for each $\varphi_\infty$ in order to take the relevant derivatives in Eqs.~(\ref{eq:charge_1})-(\ref{eq:charge_3}) and we need $\{ m_A(\bar{m}_A),\,\bar{m}_A(m_A) \}$ in order to freely switch back and forth between baryonic and gravitational mass\footnote{The gravitational mass is the one appearing in parameterized-post-Keplerian parameters that get constrained from binary pulsar experiments. Therefore, we need the baryonic mass to take the appropriate derivatives for the scalar charges, and the gravitational mass to link our results to binary pulsar experiments.}. For the two masses, we interpolate them with a simple linear method to remove any possible artifacts that arise from the interpolation itself. For $\alpha_A$ we implement different interpolation schemes depending on if spontaneous scalarization occurs. If spontaneous scalarization is absent, we simply use a cubic spline on $\alpha_A$ and this does great since the curves are smooth and generally free of any numerical anomalies. If spontaneous scalarization is present, however, then we use a cubic spline on $\log_{10} \alpha_A$ as this helps us better handle the rapid growth of the scalar field, especially when $\alpha_0 < 10^{-4}$. We find that the errors, discussed in Sec.~\ref{sec:error:masses}, are significantly smaller when we interpolate $\log_{10} \alpha_A$ instead of just $\alpha_A$. Lastly, we interpolate $\log I_A$ with a cubic spline as well, and while this may introduce some error for low masses, it better suites the data for larger masses, c.f Sec.~\ref{sec:error:masses} for a discussion.

What follows after the interpolation of the raw data is the calculation of the scalar charges $\beta_A$ and $k_A$ according to Eq.~(\ref{eq:charge_2}) and Eq.~(\ref{eq:charge_3}) respectively, making use of the various finite difference schemes in Eqs.~(\ref{eq:fd_central})-(\ref{eq:fd_backward}). At this point we are able to produce the data we have shown in our figures thus far, and we have the ability to sample our results as finely as we need to in $m_A$ in order to produce the most accurate results. However, when producing the master data files for each theory-EOS combination we do not have the luxury of over sampling in $m_A$ otherwise each individual data file would be far too large in size. Therefore, we decided to sample in the region $1 M_\odot < m_A < M_{\text{max, GR}}$, with a spacing $\Delta m_A = 0.002 M_\odot$, since this is a generous mass range in which we expect to observe pulsars.

%--------------------------------------------------------------------------------
\subsection{Properties of the master data file}

The data files we have generated contain nine columns of data and they have the structure that appears in Table~\ref{tab:data_file}. The first three columns of the data correspond to the scalar charges $\alpha_A$, $\beta_A$, and $k_A$ respectively. The fourth and fifth columns contain the values for baryonic mass and gravitational mass respectively, the latter of which falls on the grid described in the previous paragraph. Columns six and seven contain $\alpha_0$ and $\log_{10} \alpha_0$, of which the former lay on the grid described in section Sec.~\ref{sec:charges:beta1}. Column eight contains the value of $\beta_0$, which ranges from $-5 \leq \beta_0 \leq +5$ and lies on a grid with spacing $\Delta \beta_0 = 0.02$ for $\beta_0 < -4.3$ and $\Delta \beta_0 = 0.1$ for $\beta_0 > -4.3$. 

While not shown in Table~\ref{tab:data_file}, we have included 4 more columns of additional information in the master data files that some readers might find useful: the ninth column contains the  central density, the tenth contains the radius, the eleventh contains the central value of the scalar field, and the twelfth contains the surface value of the scalar field. However, we point out that in region V not all of these quantities are calculated explicitly because we make use of the aforementioned scaling relations and therefore some assumptions have been made. Because the scalar field is relatively (compared to unity) small in region V, the radius and central density can be assumed to obey the same functional relationship to the gravitational mass as those solutions found on the horizontal black dashed line appearing in Fig.~\ref{fig:parameter_space}, i.e. the set of solutions we apply the scaling relations to. One can see from the boundary condition in Eq.~\eqref{eq:BC-phiinf}, that the surface value of the scalar field, $\varphi_s$, scales directly with $\varphi_\infty$, at least to first order in $\varphi_\infty$, and thus we have made use of this in the construction of the data files. However, the central value of the scalar field cannot be assumed to obey the same relations because is is one of the free parameters the we must numerically determine through a shooting method. While it would be reasonable that this would also obey the same relations in the limit of weak scalar fields, we could not verify this from our numerical data and have therefore given them a value of ``0'' in the data files. We emphasize that the true value of the central scalar field \emph{is not} actual zero, but for the sake of providing a complete data file that can be easily interpolated we have included these null values as a place holder.

The files we have generated can be used in a variety of fields where scalar charges appear, not just binary pulsar experiments, although this is the primary target of this work. The data can be accessed though a git repository~\cite{DATA} and is structured on a uniform grid such that one can either use the data as is, or interpolate it if desired. For interpolation purposes, we have also included three separate files containing the grid points in $\{m_A,\,\alpha_0,\,\beta_0 \}$ that we used and are labeled appropriately in the git repository. We have included both a \textsc{\small Python} and \textsc{\small Mathematica} script that is capable of reading in the data, setting up the numerical grid, and interpolating the data using \textsc{\small SciPy}'s \texttt{\small RegularGridInterpolator} function for \textsc{\small Python} and \textsc{\small Mathematica}'s native \texttt{\small Interpolation} function, both of which make use of a linear order method for three-dimensional data. \/

Recall that for the gray region of parameter space in Fig.~\ref{fig:parameter_space} we were not able to find NS solutions that were of any use. As we have mentioned earlier, excluding this data from our investigation is not necessarily a shortcoming since these regions of parameter space are so heavily constrained by solar system observations and they are generally excluded from any analysis. However, for the sake of enforcing that \emph{all} data lie on a uniform grid and can therefore be interpolated with ease, we found it beneficial to include this region of parameter space in our data. To do this, we artificially give values to the three scalar charges as a sort of place holder in the data. Since these regions of parameter space are generally excluded by observations we have given the scalar charges all an unphysical value of $10^5$. We chose this value more out of convenience in an effort to force, say, an MCMC to avoid these points in parameter space because NSs do not exist and therefore the charges technically do not exist either.

%----------------------------------------------------------------
\subsection{Comparison to full data}

One questions we need to concern ourselves with is whether or not the data in these files accurately reproduces the raw data from which they came since we have had to, in some cases, undersample the data in $m_A$. Here we discuss how the data recovers the results presented in Sec.~\ref{sec:charges}, which were produced with full numerical data. The points in parameter space we consider here lie on our numerical grid and therefore should show great agreement with the full data, provided we sampled the charges finely enough in $m_A$. A more detailed discussion of the limitations of our data can be found in Appendix~\ref{sec:limitations} where we point out some known issues.

\begin{figure}[t]
	\centering
	\includegraphics[width=3.4in]{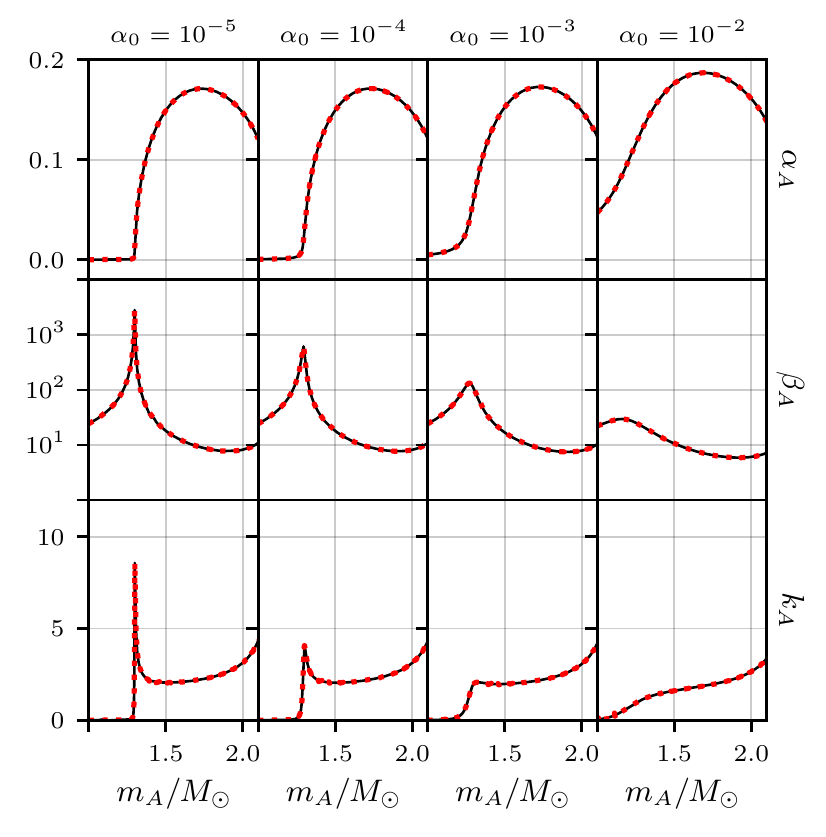}
	\caption[Numerical grid example for $m_A$]{ 
		\label{fig:error_data_file1} Comparison between the charges from our data files (dotted red) to the full raw data used to make the files (black) for points 2-5 in Table~\ref{tab:special_points}. There is excellent recovery for the full data, even in this extremely non-linear part of parameter space, suggesting that we have indeed sampled the solutions fine enough in $m_A$.
	}
\end{figure}

Figure~\ref{fig:error_data_file1} shows a comparison between some of the full data found in Fig.~\ref{fig:ss}, i.e. points labeled 2-5 in Table~\ref{tab:special_points}, and the data file for MO theory and AP3 EOS. As one can see, there is remarkable consistency between the two data sets, even in the most non-linear regions of parameter space, e.g. $\beta_0 = -5$ and $\alpha_0 = 10^{-5}$. In fact, the only real error that is noticeable is in $\beta_A$ for these parameters, and only near the sharpest part of the peak does the data files deviate from the full data 

Even deviations as large as 10\% should not have a significant effect on binary pulsar constraints, and the reason is threefold. First, the peaks in $\beta_A$ are extremely isolated in $m_A$ and there may not be a pulsar with that particular mass. Second, even if this value of $m_A$ were important, the value of $\beta_A$ is still so large that it would be immediately ruled out. Third, this error only occurs in the regions of parameter space that are already tightly constrained~\cite{Damour:1996ke,Freire:2012mg,Anderson:2016aoi,Anderson:2017phb}. The astute reader may point out that the location of the peaks in $\beta_A$ depend on the EOS and $\beta_0$, which would make it quite possible for an observed pulsar to lie on one of these peaks at some points in the $\{\alpha_0,\,\beta_0\}$ plane. While this is true, our second point above still holds and, in fact, the differences between the tabulated and full data tends to decrease as $\beta_0$ becomes less negative and $\alpha_0$ becomes larger.

%------------------------------------------------------------------------------------------------
\section{Conclusion}\label{sec:conclusion}
%------------------------------------------------------------------------------------------------

We have  extended the original work in Ref.~\cite{Damour:1996ke} and calculated the scalar charges $\{ \alpha_A ,\,\beta_A ,\,k_A \}$ for a large region of the $\{\alpha_0 ,\,\beta_0 \}$ parameter space and 11 physical equations of state, in two distinct scalar tensor theories (that of Damour-Esposito-Far\'ese and Mendes-Ortiz). We have presented the numerical schemes we implemented to complete these calculations and presented our results. Our goal was to calculate the scalar charges and tabulate them so that they could be of use in the future, particularly in the application of binary pulsar tests of gravity. We have investigated both the error of our numerical solutions, as well as the ability of certain scaling relations to reproduce full numerical results. Through this paper, the data is made fully available to the community. 

Future work that utilizes this data include tests of STTs with binary pulsars and gravitational waves. In particular, this data makes it possible to perform a Bayesian analysis on the PPK parameters of binary pulsar system. Such analysis would require the use of an MCMC in which case one would need knowledge of the scalar charges in order to compute the likelihood. Instead of calculating the scalar charges on the fly, which we have demonstrated to extremely computationally expensive, one can make use of the data file provided in this paper to significantly speed up these likelihood evaluations. We intend to perform such an investigation in future work.

%{\ny{Say something about the future of binary pulsar tests of STTs. (i) other people can now do a bayesian analysis and thus a more correct analysis of tests of STTs with binary pulsars using our data, (ii) once SKA comes online, you may detect new pulsars, and in particular BH-NS pulsars, and tests of STTs with that new data could use our data.}}

LIGO and VIRGO will inevitably detect more NS mergers in the future, some of which are likely to be NS-BH systems. Such systems are ideal for testing STTs because BHs do not develop scalar charges in these theories and therefore the emission of dipolar GWs would be maximized in these scenarios. The data provided in this paper would again be necessary for one to perform a Bayesian analysis of the data through MCMC simulations. Furthermore, if one wishes to study the constraints the future GW detectors can place on STTs, high resolution calculations of the scalar charges would be a crucial for such an analysis.

%
%{\ny{Say something about future GW tests of STTs. (i) LIGO and Virgo are going to detect more GWs, and in particular, probably NS-BH binaries. These binaries are ideal for STT tests, and the data presented here would be necessary to do MCMC by the LSC.  (ii) studies of how well 3G detectors can test STTs will require high resolution scalar charges, like those computed in this paper.}}

%------------------------------------------------------------------------------------------------

%------------------------------------------------------------------------------------------------
\acknowledgements\label{ackno}
%------------------------------------------------------------------------------------------------
We would like to thank Norbert Wex, Hector O' Silva, and Travis Robson for useful insight and discussions. DA would like to thank Paulo Freire and the Max-Planck-Institut f\"ur Radioastronomie for their hospitality during part of this work. We would also like to acknowledge the support of the Research Group at Montana State University through their High Performance Computer Cluster Hyalite. NY and DA acknowledge support from NSF grant PHY-1759615 and NASA grants NNX16AB98G and 80NSSC17M0041.

%------------------------------------------------------------------------------------------------

\appendix
%------------------------------------------------------------------------------------------------
\section{Error Analysis of Results}\label{sec:error}
%------------------------------------------------------------------------------------------------

Here we attempt to quantify the numerical error in our results. We start with an investigation of the error associated with our grid spacing in $\jrho_c$ and show that we can trust our results to within 1\% in the worst of cases for the regions of parameter space that we have explored. We then look into how reliable the scaling relations are and how much error we introduce by using these rather than populating the entire parameter space fully numerically. We then finish with a discussion of the effects of using piecewise polytrope versus their tabulated counterparts. For our analysis, we focus on the numbered points appearing in Fig.~\ref{fig:parameter_space}, which we have detailed below in Table~\ref{tab:special_points}. For each of the analyses we perform, we investigate the error at these points in parameter space in detail in an attempt to quantify our errors across the entire parameter space.

%------------------------------------------------------------------------------------------------

\begin{table}[h!]
	\centering
	\renewcommand{\arraystretch}{1.3}
	\begin{tabular*}{3.4 in}{C{0.4 in} |C{.4 in} |C{.75 in} ||C{.5 in} |C{.5 in} |C{.5 in} C{1 pt}}
		\hline
		\hline
		\# 	 & $\beta_0$ & $\alpha_0$ & $\alpha_A (\jrho_c)$ & $\beta_A (\jrho_c)$ & $k_A (\jrho_c)$ & \\ [1 pt] \hline
		1	 & 	-6		 & 	$1.44\times10^{-2}$ & --- & --- & ---  \\ [1 pt] \hline
		2	 & 	-5		 & 	$10^{-5}$ 	& $10^{-2}$  & $10^{0}$ & $10^{0}$  \\ [1 pt] \hline
		3	 & 	-5		 & 	$10^{-4}$ 	& $10^{-2}$  & $10^{0}$ & $10^{-1}$  \\ [1 pt] \hline
		4	 & 	-5		 & 	$10^{-3}$ 	& $10^{-3}$  & $10^{-2}$ & $10^{-2}$ \\ [1 pt] \hline
		5	 & 	-5		 & 	$10^{-2}$ 	& $10^{-3}$  & $10^{-3}$ & $10^{-3}$ \\ [1 pt] \hline
		6	 & 	-5		 & 	$10^{-1}$ 	& $10^{-3}$  & $10^{-1}$ & $10^{-1}$  \\ [1 pt] \hline
		7	 & 	-5		 & 	$10^{-0}$ 	& $10^{-3}$  & $10^{-1}$ & $10^{-1}$  \\ [1 pt] \hline
		8	 & 	-3		 & 	$10^{-3}$ 	& $10^{-6}$  & $10^{-5}$ & $10^{-2}$  \\ [1 pt] \hline
		9	 & 	-3		 & 	$10^{-1}$ 	& $10^{-5}$  & $10^{-2}$ & $10^{-2}$ \\ [1 pt] \hline
		10	 & 	3		 & 	$10^{-3}$ 	& $10^{-6}$  & $10^{-5}$ & $10^{-1}$\\ [1 pt] \hline
		11	 & 	3		 & 	$10^{-1}$ 	& $10^{-5}$  & $10^{-2}$ & $10^{-1}$  \\ [1 pt] \hline
		12	 & 	-4		 & 	$10^{-4}$ 	& $10^{-5}$  & $10^{-4}$ & $10^{0}$ \\ [1 pt] \hline
		13	 & 	-4		 & 	$10^{-3}$ 	& $10^{-5}$  & $10^{-4}$ & $10^{-2}$ \\ [1 pt] \hline
		14	 & 	-4		 & 	$10^{-2}$ 	& $10^{-5}$  & $10^{-5}$ & $10^{-3}$\\ [1 pt] \hline
	\end{tabular*}
	\caption{\label{tab:special_points} Numbered points appearing in Fig.~\ref{fig:parameter_space} and there associated relative errors as determined by Eq.~(\ref{eq:rel_error}). The columns with $\jrho_c$ in parentheses are the errors determined in Sec.~\ref{sec:error:masses}. Recall that Eq.~(\ref{eq:rel_error}) actually yields the percent error in the solutions, therefore $10^0$ actually represents a 1\% error, which is the highest error we find from our results.
	}
\end{table}

%------------------------------------------------------------------------------------------------
\subsection{Grid in central density}\label{sec:error:masses}
%------------------------------------------------------------------------------------------------

%\begin{figure*}[t]
%	\centering
%	\includegraphics[width=6in]{error_masses/error_masses_COSH.pdf}
%	\caption[Numerical grid example for $m_A$]{ 
%		\label{fig:error_masses} The percent error in the scalar charges $\alpha_A$, $\beta_A$, and $k_A$ for points labeled 2, 8, 10, and 13 in Table~\ref{tab:special_points}. These results are for MO theory and using AP3 for the EOS.
%	}
%\end{figure*}

In order to obtain the most accurate and precise results from our numerical calculations one would have to use an infinitely dense grid in central density $\jrho_c$ such that interpolating between points in Fig.~\ref{fig:mass_grid} leaves no room for error. However, this is not computationally feasible, and thus, by reducing the number of points in this grid we introduce numerical error that is not associated with our methods of solving the field equations. The grid we have used in the end, for each individual region of parameter space, was chosen to achieve sufficient confidence in our results for the amount of computation time needed for the calculations. While we are confident in the grids we have chosen, there is still numerical error associated with our choices and we quantify those here. 

In order to assess our errors we decided to double the number of points in our central density grid and calculate the relative errors between these solutions and the ones we have calculated using our original grids, which we have done for every point in parameter space detailed in Table~\ref{tab:special_points}. While we analyzed each of these points in detail, we only discuss the results for the points with the worst errors and give explanation for why the errors arise. For concreteness, to calculate relative percent error for any of the scalar charges, denoted by ``$\chi$'' here,  we use
\be
\dfrac{\Delta_{\text{rel}}\, \chi(\alpha_0, \beta_0, m_A)}{100} \eq \left| \dfrac{\chi_2(m_A) - \chi_1 (m_A)}{\chi_2(m_A)} \right|_{\alpha_0, \beta_0}\,\,,
\label{eq:rel_error}
\ee
where $\chi_2$ is the solution with twice as many grid points as $\chi_1$. We evaluate this for all values of $m_A$ for the specific combination of $\alpha_0$ and $\beta_0$ and quote the largest values of relative error in the columns of Table~\ref{tab:special_points} with $\jrho_c$ in parentheses. 

\begin{figure}[b]
	\centering
	\includegraphics[width=3.5in]{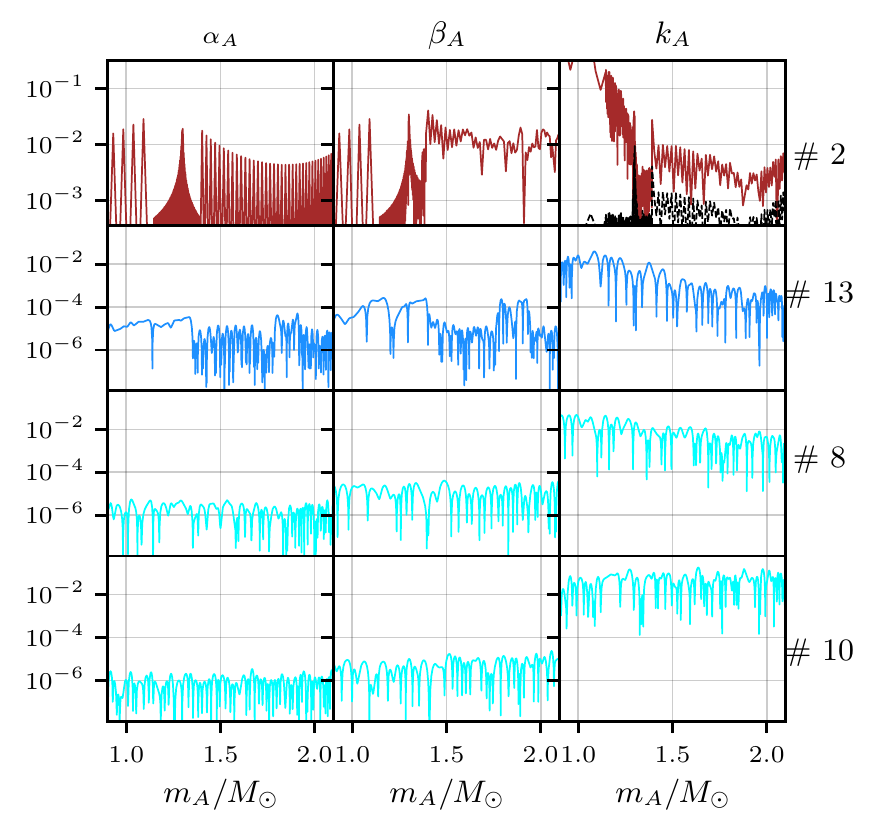}
	\caption[Numerical grid example for $m_A$]{ 
		\label{fig:error_masses} The percent error in the scalar charges $\alpha_A$, $\beta_A$, and $k_A$ for points labeled 2, 8, 10, and 13 in Table~\ref{tab:special_points}. These results are for MO theory and using AP3 for the EOS.
	}
\end{figure}

Figure~\ref{fig:error_masses} shows a representative sample of our errors, including points labeled 2, 8, 10, and 13 as they show special features worth discussing. The first row in Fig.~\ref{fig:error_masses}, showing the errors in the charges for point 2 in Table~\ref{tab:special_points}, represents what we consider to be one of the most error prone regions of parameter space. Spontaneous scalarization occurs here, and since $\alpha_0$ is so small the phase transition in the scalar field is extremely sharp. However, our results for $\alpha_A$ are good to withing 0.01\%, with some of the worst error appearing exactly when the phase transition occurs ($\sim 1.3 M_\odot$). The errors in $\beta_A$ are slightly worse as one might expect since a numerical derivative is involved, but surprisingly enough the largest error is not associated with the phase transition. The spike in the error at $\sim 1.4 M_\odot$ is related to a mishap in the interpolation of the finer grid data, and while this does not happen often, our solutions are sometimes prone to this type of error in this part of parameter space.

%\begin{figure*}[t]
%	\centering
%	\includegraphics[width=6in]{error_scaling/error_scaling.pdf}
%	\caption[Numerical grid example for $m_A$]{ 
%		\label{fig:error_scaling} The percent error in the scalar charges $\alpha_A$, $\beta_A$, and $k_A$ for points labeled 8 10 in Table~\ref{tab:special_points} when comparing scaled charges to their fully numerical counterparts. Blue curves correspond to $\alpha_0 = 10^{-2}$, red to $\alpha_0 = 10^{-4}$, and black to $\alpha_0 = 10^{-5}$. These results are for MO theory and using AP3 for the EOS. The results for DEF theory and AP3 looked nearly identical but the errors are overall slightly smaller.
%	}
%\end{figure*}

The error in $k_A$ may seem alarming at first for low masses, but this only occurs because $k_A \sim 0$ here and even tiny numerical noise in the results can generate a large relative error in the solution. This error is not a consequence of our numerical grid, but rather is an error associated with a the fact that numerical integration has finite precision and how we interpolate our results. For comparison we have also included the absolute error for $k_A$ plotted by a dashed red line in Fig.~\ref{fig:error_masses}. An obvious downfall here might be that we should have used a finer grid for the low masses, which would have most certainly increases the accuracy of our interpolations somewhat. The other issue, however, could be the method of interpolation we used. As described in Sec.~\ref{sec:data}, when calculating $k_A$ we must interpolate $\log I_A(\bar{m}_A)$ for multiple values of $\alpha_0$ in order to calculate derivatives, and we use a cubic spline to do so. Using a cubic spline on potentially noisy data like this is a good way to introduce extra error, which is what we are seeing here. However, switching to a linear order interpolation method significantly increases our errors for larger masses, precisely in the more interesting regions of parameter space where pulsar masses tend to lie. For this reason, we sacrifice precision on the lower end of masses in order to increase it elsewhere for more relevant masses.

The second, third, and fourth rows of Fig.~\ref{fig:error_masses} tell a different story than the first. For all these points in parameter space we find great agreement between solutions and therefore very small relative error. As expected, error in $\alpha_A$ are extremely small and this is due to the fact that there is no spontaneous scalarization and it is easy to extract $\alpha_A$ from the NS solutions. The errors in $\beta_A$ are slightly worse than those of $\alpha_A$ but still show exceptional agreement. As we saw with the with point 1, the errors in $k_A$ tend to be much worse than those of the other scalar charges.
%------------------------------------------------------------------------------------------------

%------------------------------------------------------------------------------------------------
\subsection{Analytic Scaling}\label{sec:error:analytic}
%------------------------------------------------------------------------------------------------
One of the greatest properties about the cyan region of parameter space in Fig.~\ref{fig:parameter_space} is that we can make use of the scaling relations presented in Sec.~\ref{sec:charges:beta3} to significantly reduce the number of NS solutions needed to explore this region. There is, however, some error associated with these scaling relations since we are, effectively, ignoring some of the non-linearities that appear in the field equations. While we expect these relations to hold in the small $\alpha_0$ regime, we need to show numerically that this is indeed the case. To investigate this particular kind of error we have decided to explore the cyan region of parameter space with the same numerical grid for the blue region, described in Sec.~\ref{sec:charges:beta1}, but only for AP3 and SLy4 EOS to get a sense of the error.

\begin{figure}[t]
	\centering
	\includegraphics[width=3.5in]{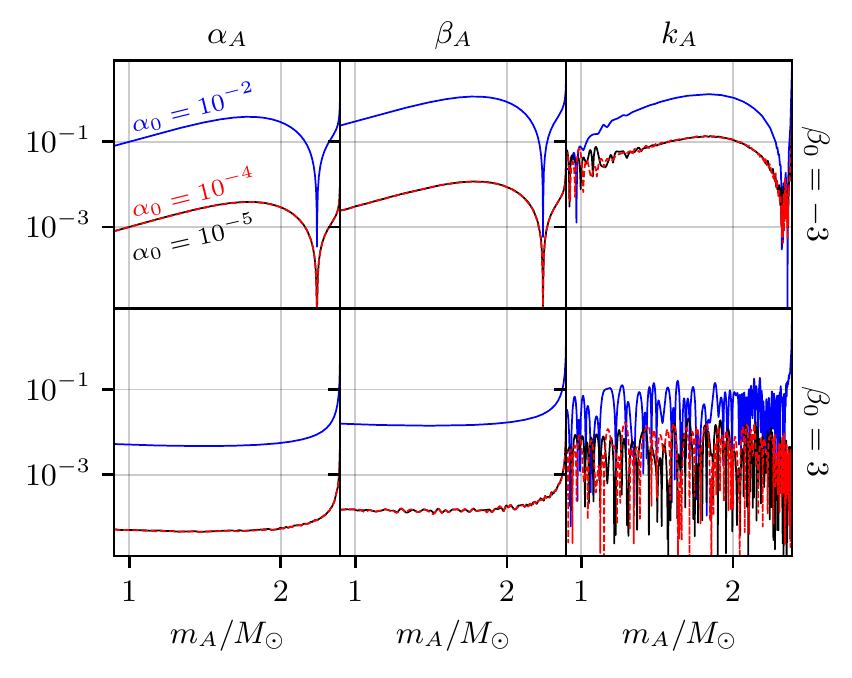}
	\caption[Numerical grid example for $m_A$]{ 
		\label{fig:error_scaling} The percent error in the scalar charges $\alpha_A$, $\beta_A$, and $k_A$ for points labeled 8 10 in Table~\ref{tab:special_points} when comparing scaled charges to their fully numerical counterparts. Blue curves correspond to $\alpha_0 = 10^{-2}$, red to $\alpha_0 = 10^{-4}$, and black to $\alpha_0 = 10^{-5}$. These results are for MO theory and using AP3 for the EOS. The results for DEF theory and AP3 looked nearly identical but the errors are overall slightly smaller.
	}
\end{figure}

To make use of the scaling relations, we simply calculate all the scalar charges on the dashed horizontal lines in Fig.~\ref{fig:parameter_space} and use Eq.~(\ref{eq:charge1_scale}) to find the function $f(\beta_0, m_A)$. Once we have found $f(\beta_0, m_A)$, then $\alpha_A$ can be solved for all other $\alpha_0 < 10^{-2}$ by substituting $f$ and the new $\alpha_0$ back into Eq.~(\ref{eq:charge1_scale}). Since we have already calculated the actual, unscaled scalar charges for the AP3 and SLy4 EOSs can can compare the scaled versions to the full numerically solved ones.

Figure~\ref{fig:error_scaling} shows some of the results from our error analysis for $\beta_0 = \pm 3$, where the scaling relations are expected to hold. We see that the relations have the most error for solutions with $\alpha_0 = 10^{-2}$ (blue curves) but are never more than 1\%. For all values of $\alpha_0 < 10^{-2}$ the error continues to decrease and would presumably go to zero in the limit that $\alpha_0 \rightarrow 0$ if we had infinite precision in our numerics. In general, we find errors of the same order of magnitude as the ones presented in Fig.~\ref{fig:error_scaling} across the entire cyan region of parameter in Fig.~\ref{fig:parameter_space}, with the errors being slightly larger in MO theory than in DEF theory. One notices that the errors for $\alpha_A$ and $\beta_A$ look very similar, and they should be this way according to Eq.~(\ref{eq:charge2_scale3}) since $\beta_A$ is a derivative of $\alpha_A$. The inertial charge $k_A$ also has a similar structure to the other charges but is polluted with more numerical noise, which is extremely evident for $\beta_0 = +3$. The distinct difference in the function form of the errors between $\beta_0 = -3$ and $\beta_0 = +3$ can be attributed to the fact that the effects of the non-linearities in the scalar field are more prominent for negative values of $\beta_0$.

%------------------------------------------------------------------------------------------------

%------------------------------------------------------------------------------------------------
\subsection{Piecewise Polytropes}\label{sec:error:peicewise}
%------------------------------------------------------------------------------------------------
The piecewise polytropic approximation to tabulated EOSs in Ref.~\cite{Read:2008iy} allows one to considerable speed up numerical calculations involving NSs because of their analytic nature. The data files we provide were all generated using the full tabulated data, but for others who wish to perform future calculation, it would be nice to have an idea how these approximations affect the final results. In Fig.~\ref{fig:def96_comp} we briefly compared how charges calculated using the piecewise polytropes compare to those calculated with the full tabulated EOS, and one can see that aside from slight apparent shifts in the masses, the curves seem nearly identical. We have verified, using the points in Table~\ref{tab:special_points} as a representative sample, that this is indeed consistent across the entire parameter space. The similarity between the charges should not come as a surprise considering how well the piecewise polytropes match the actual data, c.f Table III in Ref.~\cite{Read:2008iy}. For all of the equations of state we consider here, the residuals in Ref.~\cite{Read:2008iy} for both the mass and moment of inertia of the NSs are less than 2\%, which are the two most important quantities when calculating the scalar charges. Having such small deviations between tabulated and approximated EOSs leads to very small deviations when calculating $\alpha_A$, $\beta_A$, and $k_A$ from Eqs.~(\ref{eq:charge_1})-(\ref{eq:charge_3}).

%------------------------------------------------------------------------------------------------

\section{Limitations of Data Files}\label{sec:limitations}

We have discovered a few limitations of the data we have provide and these are discussed in this appendix. We should point out, however, that these complications arise when one wishes to interpolate the data files we have generated, in order to determine that charges for points that \emph{do not} lie on our numerical grid. As far as we are able to tell, the data files are able to reproduce the full numerical data used to make the files to great level of accuracy (see Sec.~\ref{sec:data}) for all points that lie on our grid.

\begin{figure}[h]
	\centering
	\includegraphics[width=3.5in]{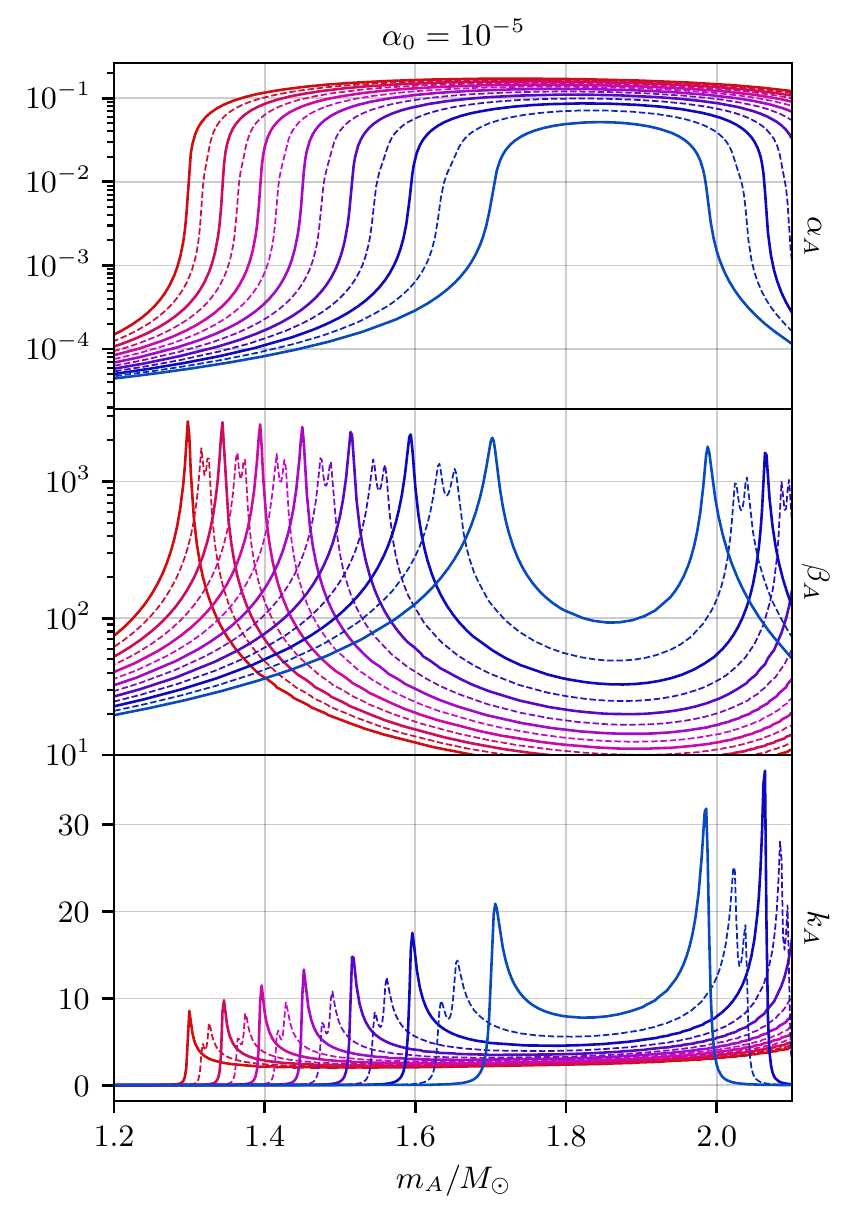}
	\caption[Numerical grid example for $m_A$]{ 
		\label{fig:error_data_file2} Comparison between the charges for points that lie on our $\{ \alpha_0 ,\, \beta_0 \}$ numerical grid (solid lines) and points the we have to interpolate to find (dashed lines). Similar to Fig~\ref{fig:ss}, we use a constant value of $\alpha_0 = 10^-5$ and $-5.0 \leq \beta_0 \leq -4.4$, ranging in color from most red to most blue respectively. One notices that the interpolated point acquire a double peaked feature, the details of which are described in the text.
	}
\end{figure}

Figure~\ref{fig:error_data_file2} shows how the interpolation of the data behaves for points that lie on (solid) and off (dashed) numerical grid we have established for $\alpha_0$ and $\beta_0$. One notices that the solid curves appear as expected, according to the results presented in Figs.~\ref{fig:def96_comp} and \ref{fig:ss}; $\alpha_A$ experiences a smooth rapid growth while $\beta_A$ and $k_A$ have peaks near $m_\crit$ for which spontaneous scalarization turns ``on'' and ``off''. However, the dashed curves, which lie off the numerical grid, have somewhat significantly different features. The curves for $\alpha_A$ appear to be good, but one notices that the dashed curves develop slight instantaneous ``discontinuities'' in the slope during the growth of the scalar field. As a result of this apparent discontinuities, $\beta_A$ develops a double peak near the critical mass at which the phase transition occurs. Likewise, a very similar feature develops in $k_A$ in which the single peak that is present in the solid curves turns into two peaks that are not as large in magnitude.

The issue we are seeing in Fig.~\ref{fig:error_data_file2} is a direct consequence of the nature of the scalar charges, particularly the presence of the phase transition, and trying to interpolate them. Consider $\beta_A$ for example, in which case we expect the magnitude of the peak at the critical mass to decrease as $\beta_0$ becomes less negative, cf. the plots of $\beta_A$ in Fig.~\ref{fig:ss}. We also expect the critical mass to shift to higher masses as we allow $\beta_0$ to become less negative. Therefore, there are multiple features changing in the solutions with the variation of just a single parameter, in this case $\beta_0$. Because of this dependence, the linear interpolator has issues when it tries to interpolate between these peaks because the algorithm considers changes in $\beta_0$ and $m_A$ at the same time, which physically are in effect co-dependent on each other in a non-linear way. As it might be expected, it is hard to interpolate any function with extremely sharp features like we have here and taking the log of the data does not seem to improve anything in this situation. 

The features described thus far are most prominent in the $\alpha_0 \leq 10^{-5}$ case since the peaks are the most narrow here. However, this artifact of the interpolation does arise for other values of $\alpha_0$ when spontaneous scalarization, but since the effects are less localized, i.e. the phase transition is more smooth for larger values of $\alpha_0$, the discrepancy is less severe. Once out of the regime of spontaneous scalarization these artifacts no longer appear to be present and the interpolation of the data does an excellent job of giving us information between our grid points.

\bibliography{thesis}
\end{document}